\documentclass[prd,aps,amsfonts,eqsecnum,nofootinbib,notitlepage,longbibliography,10pt,superscriptaddress]{revtex4-1}

\usepackage{graphicx}
\usepackage{tabularx}
\usepackage{braket}
\usepackage{amssymb,mathtools}
\usepackage{amsmath}
\usepackage{bm}
\usepackage[driverfallback=dvipdfm]{hyperref}
\hypersetup{colorlinks=true,breaklinks,urlcolor=blue,linkcolor=blue,citecolor=blue}

\usepackage{xcolor}

\newcommand{\comment}[1]{}
\definecolor{ginger}{rgb}{0.69, 0.4, 0.0}

\newcommand{\lr}[1]{\left( #1\right)}
\newcommand{\mlr}[1]{\left[ #1\right]}

\newcommand{\alr}[1]{\left\langle #1\right\rangle}
\newcommand{\norm}[1]{\left\lVert#1\right\rVert}
\newcommand{\abs}[1]{\left\lvert#1\right\rvert}

\newcommand{\ii}{\mathrm{i}}
\newcommand{\ee}{\mathrm{e}}
\newcommand{\dd}{\mathrm{d}}
\newcommand{\tr}[1]{\mathrm{Tr}\lr{#1}}
\newcommand{\trace}{\mathrm{Tr}}

\newcommand{\diam}{\mathrm{diam}}
\newcommand{\where}{\quad {\rm where}\quad}
\newcommand{\order}{\mathrm{O}}

\newcommand{\CC}{\mathcal{C}}
\newcommand{\cR}{\mathcal{R}}

\newcommand{\cL}{\mathcal{L}}

\newcommand{\DD}{{\mathcal{D}}}
\newcommand{\poly}[1]{\mathrm{poly}\lr{#1} }

\newcommand{\OO}{\mathcal{O}}
\newcommand{\cU}{\mathcal{U}}
\newcommand{\cH}{\mathcal{H}}

\newcommand{\kpsi}{\ket{\psi}}
\newcommand{\kpi}{\ket{\sqrt{\pi}}}

\newcommand{\bz}{{\bm{z}}}

\newcommand{\kplus}{\ket{\bm{+}}}
\newcommand{\pushright}[1]{\ifmeasuring@#1\else\omit\hfill$\displaystyle{#1}$\fi\ignorespaces}
\newcommand{\kett}[1]{\left|#1\right\rangle\mkern-9mu\left.\vphantom{#1}\right\rangle}
\newcommand{\bbra}[1]{\left\langle\vphantom{#1}\right.\mkern-9mu\left\langle#1\right|%
}
\newcommand{\mytag}[1]{\tag*{\llap{$\lr{#1}$}}}

\makeatletter
\renewcommand{\p@subsection}{}
\renewcommand{\p@subsubsection}{}
\makeatother

\usepackage{amsthm}
\newtheorem{thm}{Theorem}
\newtheorem{cor}[thm]{Corollary}
\newtheorem{lem}[thm]{Lemma}

\newtheorem{prop}[thm]{Proposition}
\newtheorem{eg}[thm]{Example}

\newtheorem{defn}[thm]{Definition}

\begin{document} 

\title{Fast mixing and thermal phases from strong decay of correlations}
\author{Chao Yin}\email{chaoyin@stanford.edu}
\affiliation{Department of Physics, Stanford University, Stanford, California 94305, USA}

\author{Vedika Khemani}\email{vkhemani@stanford.edu}
\affiliation{Department of Physics, Stanford University, Stanford, California 94305, USA}

\date{\today}

\begin{abstract}
We show that a strong form of correlation decay in quantum Gibbs states has far-reaching dynamical consequences. For a smooth path of geometrically local commuting Hamiltonians whose Gibbs states satisfy strong decay of correlations, we construct $\mathrm{O}(1)$-time quasi-local Lindbladian evolutions exactly connecting these states in both directions, together with $\mathrm{O}(1)$-time quasi-local Hamiltonian evolutions exactly connecting their canonical purifications. These quasi-local connections strengthen previously known quantum-channel circuit constructions based on the decay of connected correlation functions [R.~Ma, V.~Khemani, and S.~Sang, PRX Quantum \textbf{7}, 033052 (2026)], which approximately connect states within the same phase but have depths that grow logarithmically with system size. Furthermore, if the Gibbs-state path contains a product state, we prove that the single-site Glauber dynamics has a $\Omega(1)$ spectral gap independent of the system size throughout the path. This establishes Glauber-dynamics fast mixing for models that may violate the strong spatial mixing condition used in prior proofs, including
(\emph{1}) the $q$-state Potts model throughout the high-temperature phase for any $q$ and any finite dimension; (\emph{2}) the Ising model with a longitudinal field at low temperature, with frustrating perturbations that destroy monotonicity of the dynamics; and (\emph{3}) the three-dimensional $\mathbb{Z}_2$ lattice gauge theory at low temperatures comparing to a field perturbation, even though the model supports classical memories at intermediate temperatures. Intriguingly, we establish fast mixing of these classical Markov chains using tools from quantum physics like Lieb-Robinson bounds. 
\end{abstract}

\maketitle
\tableofcontents

\section{Introduction}
The mixing time of Markov chains is a longstanding research topic at the intersection of mathematics, physics, and computer science \cite{levin_markovchains,MonteCarlo_rev09,MonteCarlo_Bayesian90,MonteCarlo_compugeometry91,MonteCarlo_permanent04}. Of particular interest are local thermalization dynamics for statistical-mechanical models, whose fast mixing enables efficient Monte Carlo simulations \cite{Metropolis53,Glauber63,MonteCarlo_SU2gauge80,MonteCarlo_3dclaHeissen91}. On the other hand, the absence of fast mixing (i.e. slow mixing) could enable robust information storage and provide a route to thermal quantum memories that operate without active error correction \cite{q_memory02,qmemory_rev16,3dqmemory26}. Since standard local thermalization dynamics is determined by the Gibbs state, a central question is \emph{which properties of the state guarantee fast mixing}. Although several general criteria have been developed \cite{DobrushinShlosman_CompleteAnalyticity87,StroockZegarlinski_logSob92,Martinelli99,SSM_rev04,spectralIndependence20,spectralIndependence21}, their applicability remains limited, leaving basic questions open even for the classical Potts model \cite{Potts_52,Potts_randcluster_mixing24}. In this paper, we prove a sufficient condition for fast mixing of commuting quantum Gibbs states based on a strong form of correlation decay in the Gibbs state, resolving such questions for several fundamental models. Intriguingly, even for classical models, our proof employs tools from quantum dynamics, including Lieb-Robinson bounds \cite{LRB72,LRB_rev10,LRB_review}, to address these mixing problems.

Our mixing proof is naturally motivated by an ``adiabatic-path'' description of phases of matter for thermal Gibbs states. Thermal phases of matter are conventionally characterized by two  mathematical approaches. The Dobrushin--Lanford--Ruelle (DLR) framework examines the possible family of infinite-volume Gibbs states obtained from different faraway boundary conditions \cite{Dobruschin_state68,Dobruschin_state69,statmech_book17}. However, the uniqueness of the DLR state is believed to be insufficient to distinguish phases without a local order parameter, such as
the deconfined and trivial phases of a gauge theory \cite{3dgauge_wegner71,3dgauge_slowmix26,3dgauge_highT_unique26}. 
A second approach characterizes thermal phases through the analyticity of the free energy \cite{leeyangzero,Fisherzero65}, yet it fails to capture nontrivial phases in local models on bounded degree graphs \cite{Eggarter, LDPC_glass_montanari06,dynPT_nothermo_Montanari06, rakovszky2023physicsgoodldpccodes, LDPC_memory_trivialF25, placke_topological_2024}. 
In contrast, for gapped ground states there is a universal characterization of two states being in the same phase, namely the existence of an $\order(1)$-time quasi-local unitary connecting the states \cite{quasiadiab05,gap_phase_unitary10,gap_phase_unitary11,quasiadiab12} (i.e. the continuous-time analogue of a shallow local unitary circuit). This suggests a corresponding definition of \emph{thermal phases based on quasi-local connections} that may avoid the difficulties above. Indeed, related ideas have been proposed in different forms in Refs.~\cite{sourcery_mixedstate16,thermal_circuit26}; see also \cite{thermal_prepare_cor_BrandaoKastoryano19} regarding efficient preparation of thermal states. However, Ref.~\cite{sourcery_mixedstate16} does not provide a general construction of such connections beyond specific settings; although Ref.~\cite{thermal_circuit26} treats general paths of Gibbs states satisfying a correlation-decay condition, its construction is weaker than the ground state case in that the states are only approximately connected, with locality of the evolution growing polylogarithmically with the system size. In this work, we identify a strong form of correlation decay that guarantees exact $\order(1)$-time quasi-local connections along the path, providing a first step toward a general classification of thermal phases analogous to that of gapped ground states.

\subsection{Quasi-local connection within a phase}

We now formulate the problems more precisely and give an overview of our main results.

We focus on local models in finite spatial dimensions, and consider two notions of quasi-local connection between two Gibbs states $\pi_0,\pi_1$ in the same system. The first is a two-way quasi-local channel connection:
\begin{equation}\label{eq:2way}
    \pi_1 = \mathcal{T} \ee^{\int^1_0 \cL^{1\leftarrow 0}(s)\dd s} [\pi_0], \qquad \pi_0 = \widetilde{\mathcal{T}}\ee^{-\int^0_1 \cL^{0\leftarrow 1}(s)\dd s} [\pi_1]
\end{equation}
where $\mathcal{T},\widetilde{\mathcal{T}}$ denotes time ordering and anti-time ordering, and both Lindbladians $\cL^{1\leftarrow 0}(s),\cL^{0\leftarrow 1}(s)$ are quasi-local: each is an extensive sum of terms whose norms decay sufficiently fast with the diameters of their supports. In the examples considered later, this decay is exponential. Equation \eqref{eq:2way} is a continuous-time version of the two-way local-channel circuit criterion used to define mixed-state phases \cite{thermal_topo_order11,dissip_generate_TO14,mixedstatephase_dissip19,mixedstatephase_QEC24,mixedstatephase_markovlength25,mixedstatephase_2dclassify25,mixedstatephase_localreversible26}: A shallow circuit with strictly local gates cannot connect states within the same phase exactly in general due to nonzero correlations at arbitrarily large separations. Requiring transformations in both directions is also essential, since on-site completely depolarizing channels map every state to the trivial maximally mixed state.

The second notion of quasi-local connection concerns the canonical purification of a state $\pi$, also called its thermofield double,
\begin{equation}\label{eq:purification}
    \kett{\sqrt{\pi}}:=\lr{\sqrt{\pi}\otimes I}\kett{\Omega},
\end{equation}
where
\begin{equation}\label{eq:Omega=}
\kett{\Omega}:=\sum_{\bz}\ket{\bz}\otimes\ket{\bz}
\end{equation}
is the unnormalized product of maximally entangled pairs between the physical system and an isomorphic auxiliary system, and $\ket{\bz}$ are the computational basis states. The state in \eqref{eq:purification} is normalized because $\trace(\pi)=1$. Two canonical purifications are in the same pure-state phase if they can be connected by a quasi-local unitary:
\begin{equation}\label{eq:pi=Upi}
    \kett{\sqrt{\pi_1}} = \mathcal{T}\ee^{-\ii\int^1_0 W(s)\dd s} \kett{\sqrt{\pi_0}}
\end{equation}
where $W(s)$ is a quasi-local Hamiltonian on the doubled system. This connection between Gibbs states is closely related to the mixed-state $s$-sourcery introduced in Ref.~\cite{sourcery_mixedstate16}. 

The two notions of quasi-local connection are both natural extensions of the quasi-local unitary connection for pure-state phases. They may be deeply related: For gate-based circuits, locally reversible channel circuits induce approximate local unitary connections between the corresponding canonical purifications \cite{mixedphases_purification26}.
Our first main result establishes both notions above for commuting Gibbs states where Hamiltonian terms commute with one another, under a general correlation-decay condition.

\begin{thm}[informal]\label{thm:first}
    If a smooth path of commuting Gibbs states $\pi_s$ ($s\in[0,1]$) satisfy strong decay of correlations, then they are connected by both two-way quasi-local channels \eqref{eq:2way} and quasi-local unitaries for their canonical purifications \eqref{eq:pi=Upi}.
\end{thm}

\begin{figure}
    \centering
    \includegraphics[width=0.8\linewidth]{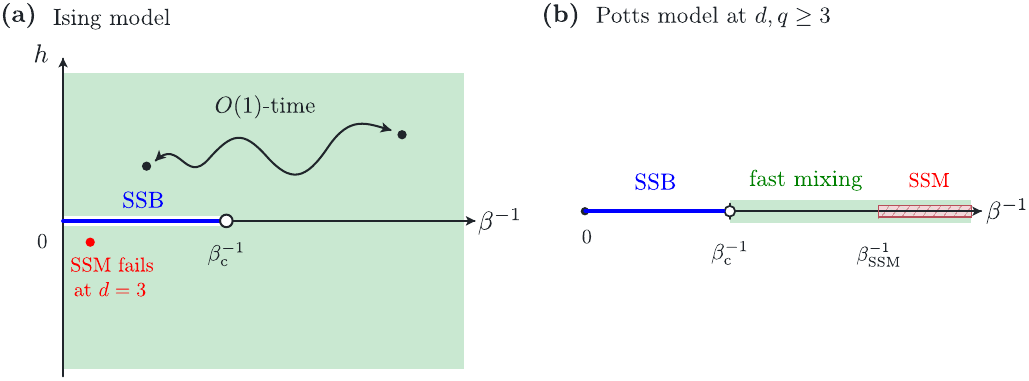}
    \caption{Phase diagrams of (a) the Ising model with a longitudinal field $h$ at temperature $\beta^{-1}$ and (b) the $q$-Potts models. $d$ is the spatial dimension. Our results apply to the green regions, where we obtain $\order(1)$-time quasi-local connections \eqref{eq:2way},\eqref{eq:pi=Upi} and a constant spectral gap for the single-site Glauber dynamics. In particular, this is proven for the Potts model in the whole high-temperature phase, where strong spatial mixing (SSM) is expected to fail at intermediate temperatures \cite{Potts_randcluster_mixing24}. SSM is also expected to fail for the $d=3$ Ising model at certain points in the green region, but its constant spectral gap is known via monotonicity of the system. Our general results apply to other models as we discuss.}
    \label{fig:Potts}
\end{figure}

Note that the quasi-local channels constructed are automatically locally reversible \cite{mixedstatephase_localreversible26} in the sense that $[\cL^{1\leftarrow 0}(s)+\cL^{0\leftarrow 1}(s)]\pi_s=0$ for any intermediate state $\pi_s$, i.e. there is a local Lindbladian $\cL^{0\leftarrow 1}(s)$ that exactly reverses the action of $\cL^{1\leftarrow 0}(s)$.
Roughly speaking, strong decay of correlations means that the state in a local region changes little when an arbitrarily large, distant region is conditioned on any configuration. This is stronger than the usual decay of connected correlation functions because the conditioning may involve an atypical configuration in the Gibbs distribution. 
Under this stronger condition, Theorem~\ref{thm:first} gives exact $\order(1)$-time evolutions that strengthen the approximate connections of Ref.~\cite{thermal_circuit26}, and unifies with the pure-state notion of \cite{sourcery_mixedstate16}.
As examples, Theorem~\ref{thm:first} applies throughout the trivial phases of the ferromagnetic Potts model and the Ising model with a longitudinal field, as shown in Fig.~\ref{fig:Potts}, and we will discuss other examples afterwards.

\subsection{Fast mixing and the spectral gap of Glauber dynamics}

A canonical model of thermalization in statistical mechanics is Glauber dynamics, which updates one site at a time according to the states of its neighbors \cite{Glauber63}. A central question is which properties of the thermal state imply a constant system-size-independent lower bound on the spectral gap of this dynamics, which we refer to as a constant spectral gap for short. The inverse of the spectral gap then yields a constant relaxation time; see precise definitions in Section \ref{sec:prelim}. A constant spectral gap implies fast mixing of the dynamics, meaning convergence to the thermal state in time at most polynomial in the system size. Although fast mixing includes Glauber dynamics at critical points taking truly polynomial time \cite{2dIsing_criticalslow12}, systems with a constant spectral gap usually obeys logarithmic mixing timescales, i.e. \emph{rapid mixing}, the fastest possible timescales one can expect.

The Dobrushin criterion provides a sufficient condition for a constant spectral gap, but it is typically restricted to regimes of weak dependence among the sites, such as sufficiently high temperature or large field \cite{Dobrushin_condition70}. Thermalization in one dimension is also fast in general \cite{rapidmix_1dcommuting23,rapidmix_1dentropy24,rapidmix_1dlogSob26,rapidmix_1d_Kim26,rapidmix_1d_Chen26}. A standard, more general correlation-decay criterion is strong spatial mixing (SSM) \cite{rapidmix_general94,Martinelli99,SSM_rev04}; related spatial-mixing criteria have also been developed for quantum Gibbs samplers \cite{Gibbssamp_commuting16,logSobolev_commuting20,fastmix_qdouble23,fastmix_cordecay_Lucia25,rapidmix_2local25,thermal_prepare_matrixCMI25,logSobolev_CSS26}. Roughly speaking, SSM requires the influence of a local change in boundary conditions to decay with distance even when the remaining boundary sites are fixed to arbitrary values. In particular, this rules out correlations that propagate over long distances along a boundary between fixed and unfixed sites, where the unfixed sites are in the conditioned Gibbs state. SSM can therefore be stronger than necessary for a constant spectral gap under standard boundary conditions, such as periodic or free ones \cite{Martinelli99}.
Indeed, the three-dimensional Ising model in a nonzero field may violate SSM in the parameter region shown in Fig.~\ref{fig:Potts}, while it is rapid mixing under periodic boundary conditions \cite{rapidmix_attractive94,Martinelli99} utilizing the monotonicity property of the system. Since monotonicity can be destroyed by weak perturbations that preserve the thermal phase, such as antiferromagnetic next-nearest-neighbor couplings, it is natural to expect it to be a feature of the proof rather than a physical requirement for a constant spectral gap. Stability of rapid mixing has been analyzed \cite{rapidmix_stability15,rapidmix_stability26} but does not resolve the question.

This raises a central question for general models: which conditions that are potentially weaker than SSM still imply a constant spectral gap for Glauber dynamics? Our second main result addresses this question from the thermal phase perspective based on quasi-local connections. 

\begin{thm}[informal]\label{thm2}
    If a smooth path of commuting Gibbs states $\pi_s$ ($s\in[0,1]$) satisfy strong decay of correlations with $\pi_0$ being a product state, then the Glauber dynamics for $\pi_1$ has a constant spectral gap.
\end{thm}

The strong decay of correlations condition is weaker than SSM, because it allows for correlations propagating along fixed boundaries.
As an application, Theorem \ref{thm2} resolves an open question \cite{Potts_randcluster_mixing24} by establishing fast mixing for the single-site Glauber dynamics of the $q$-Potts model throughout its high-temperature phase, for any $q$ and any fixed finite spatial dimension (see Fig.~\ref{fig:Potts}). It also applies to perturbed Ising models away from monotonicity, suggesting fast mixing as a property of the phase, robust to changes in the microscopic Hamiltonian.
Furthermore, it also proves fast mixing
in the low-temperature Higgs regime in the classical 3d $\mathbb{Z}_2$ gauge theory \cite{3dgauge_wegner71} with a field; together with the slow mixing regime at higher temperature \cite{3dgauge_hastings19,3dgauge_slowmix26}, this confirms the memory phase as entropy-driven which only exists at intermediate temperatures. Beyond these new results, all known finite-dimensional commuting models with known constant spectral gaps are captured by Theorem \ref{thm2} to the best of our knowledge, where a family of examples are studied in our companion work \cite{noredundancy26}.

\subsection{Outline of the paper}

Section~\ref{sec:prelim} introduces the mathematical settings of the commuting Hamiltonians we consider, together with the thermalization dynamics.
We then develop our results first for classical Gibbs states, where the correlation-decay conditions and their dynamical consequences are more transparent.
Section~\ref{sec:SDOC} defines strong decay of correlations and compares it with other conditions including strong spatial mixing.
Section~\ref{sec:connection} constructs quasi-local connections between Gibbs states by manipulating the state change $\partial_s \pi$ using strong decay of correlations, where Theorem \ref{thm:cohe} proves the classical case of Theorem \ref{thm:first}.
Section~\ref{sec:fast} proves the classical case of Theorem \ref{thm:first} in Corollary \ref{cor:SDOC_gap}, by combining strong decay of correlations with Lieb--Robinson bounds and a local gap property of the Glauber dynamics.
Section~\ref{sec:applications} applies these results to various classical models.
Section~\ref{sec:commuting} extends the framework to commuting quantum Hamiltonians.
Finally, Section~\ref{sec:outlook} discusses open questions and possible extensions.

\section{Preliminaries}\label{sec:prelim}

Consider $N$ qudits on the vertices $\Lambda$ of a $d$-dimensional square lattice with periodic boundary condition. Each qudit has dimension $\le q$. We consider local Hamiltonians of form \begin{equation}\label{eq:H=sum_s}
    H=\sum_{u\in \Lambda} H_u
\end{equation}
where each $H_u$ is a local term acting on a hypercube of $3^d$ qudits centered at $u$\footnote{Hamiltonians of larger range can be put to this form by combining nearby vertices and enlarging the single-site Hilbert space.}.
Beyond edges in the original square lattice, we also connect two vertices $u,v$ by an edge if there is a local term $H_{u'}$ acting on both of them. We view such two vertices as neighbors with distance $\mathrm{dist}(u,v)=1$. More generally, we define the distance function $\mathrm{dist}(u,v)$ as the number of edges in the shortest path connecting $u,v$ in the graph $G=(\Lambda,E)$ with $E$ the set of edges above. From the spatial dimension $d$, there exists a constant $c_d\ge 1$ such that \begin{equation}\label{eq:cd}
    \abs{\{v: \mathrm{dist}(u,v)\le r+1\} }\le c_d (r+1)^d, \qquad \abs{\{v: \mathrm{dist}(u,v)= r\} }\le c_d (r+1)^{d-1}
\end{equation}
for all $r\in \mathbb{Z}_{\ge 0}$. Here $\abs{\cdot}$ is the cardinality of the set $\cdot$.

For any set $S\subset \Lambda$, define its diameter \begin{equation}
    \diam(S):=\max_{u,v\in S}\mathrm{dist}(u,v).
\end{equation}
The graph distance $\mathrm{dist}(u,S):=\min_{v\in S}\mathrm{dist}(u,v)$.
We say $S$ is connected if it cannot be separated $S=S_1\cup S_2$ such that $S_1,S_2$ are disconnected in the graph $G$.
Let $S^{\rm c}=\Lambda\setminus S$ be the complement of $S$ in $\Lambda$, and \begin{equation}
    \partial S:=\{u\in S^{\rm c}:\mathrm{dist}(u,S)=1\}
\end{equation} 
be its neighbors. The union with its neighbors is denoted by $\overline{S}:=S\sqcup \partial S$. Sometimes we use shorthand notations $AB:=A\sqcup B$ for disjoint union of sets, and $\partial u:=\partial\{u\},\overline{u}:=\overline{\{u\}}$ for the case of a single site $u\in \Lambda$. We use $\OO_S$ to denote an operator supported in $S$, i.e. it is identity outside of $S$. 


In this paper, we require $H$ to be commuting in that \begin{equation}
    [H_u,H_v]=0
\end{equation}
for all $u,v\in \Lambda$. The Hamiltonian is normalized such that \begin{equation}\label{eq:Hu<1}
    \norm{H_u} \le \beta
\end{equation}
i.e. we absorb the inverse temperature into the Hamiltonian. Here $\norm{\cdot}$ is the operator norm, and we will use $\norm{\cdot}_1$ for the Schatten $1$-norm, i.e. the sum of singular values.
The Gibbs state is 
\begin{equation}\label{eq:pi_global}
    \pi = \frac{\ee^{-H}}{\tr{\ee^{-H}} }
\end{equation}
with expectation value of operator $\OO$ denoted by \begin{equation}
    \alr{\OO}_\pi := \tr{\pi\OO}.
\end{equation}

Since $H$ is commuting and has finite interaction range, the Gibbs state $\pi$ is a Markov network \cite{thermal_markov12}: For any $A,B,C\subset \Lambda$ with $ABC=\Lambda$ and $\mathrm{dist}(A,C)>1$, \begin{equation}\label{eq:Ipi=0}
    I_\pi(A:C|B)=0
\end{equation}
for conditional mutual information \begin{equation}
    I_\pi(A:C|B):=\alr{\log\pi+\log\pi_B-\log\pi_{AB}-\log\pi_{BC}}_\pi
\end{equation}
where $\pi_B$ for example is the reduced density matrix of $\pi$ in $B$, and $\log \pi_B$ is understood as $(\log \pi_B)\otimes I_{B^{\rm c}}$ with identity operator outside of $B$.

\subsection{Dynamics and mixing time}

We consider local Lindbladians of form \begin{equation}\label{eq:L=Lu}
    \cL=\sum_{u\in \Lambda} \cL_u
\end{equation} 
where each $\cL_u$ is a superoperator acting on $\overline{u}$ (i.e. $u$ and its neighbors), such that it generates a quantum channel $\ee^{t\cL}$ for any time $t\ge 0$.

A canonical choice of $\cL$ that preserves the Gibbs state \begin{equation}\label{eq:Lupi=0}
    \cL_u[\pi] = 0 \qquad (\forall u\in \Lambda)
\end{equation}
is the single-site heat-bath dynamics (See e.g. \cite{Gibbssamp_commuting16}): \begin{equation}\label{eq:glauber=}
    \cL_u[\rho]=\sqrt{\pi_{\overline u}}\,\pi_{\partial u}^{-1/2}
    \lr{I_u\otimes\trace_u\rho}\pi_{\partial u}^{-1/2}\sqrt{\pi_{\overline u}}-\rho.
\end{equation}
We will refer to this particular dynamics as Glauber dynamics for short, borrowing the name from the classical case.
Heuristically, at each discrete time step $t'\in \mathbb{Z}_{>0}$ of this dynamics, a random vertex $u\in\Lambda$ is picked and resampled based on the state on its neighbors $\partial u$ from the Markov property \eqref{eq:Ipi=0}. The continuous time corresponds to $t=t'/N$. The Glauber dynamics \eqref{eq:glauber=} satisfies \emph{detailed balance} with respect to $\pi$: 
\begin{equation}\label{eq:detail_ba}
    \alr{\OO,\cL^\dagger[\OO']}_\pi = \alr{\cL^\dagger[\OO],\OO'}_\pi
\end{equation}
for any operators $\OO,\OO'$,
where $\cL^\dagger$ is the conjugation of $\cL$ with respect to the Hilbert-Schmidt inner product: \begin{equation}
    \tr{\rho\cL^\dagger[\OO]}=\tr{\cL[\rho]\OO}
\end{equation}
and $\alr{\OO,\OO'}_\pi:=\tr{\OO^\dagger \sqrt{\pi}\OO' \sqrt{\pi}}$.

We define the mixing time $t_{\rm mix}$ of $\cL$ with steady state $\pi$ as the smallest value of $t\ge 0$ such that \begin{equation}\label{eq:mix<1/4}
    \norm{\ee^{t\cL}[\rho]-\pi}_1\le \frac{1}{4}
\end{equation}
for any initial state $\rho$. Note that a more general $t_{\rm mix}(\epsilon)$ can be defined by replacing $1/4$ in \eqref{eq:mix<1/4} with a small parameter $\epsilon$, which only leads to an extra term $\log \frac{1}{\epsilon}$ in the mixing time \cite{levin_markovchains}. We say the dynamics is \emph{rapid mixing}, if \begin{equation}\label{eq:rapid}
    t_{\rm mix}\le c_{\rm rapid}\log N
\end{equation}
for a constant $c_{\rm rapid}$ independent of $N$, for sufficiently large system size $N$. On the other hand, we say the dynamics is \emph{fast mixing}, if \begin{equation}
    t_{\rm mix}\le N^{c_{\rm fast}}
\end{equation}
for a constant $c_{\rm fast}$ independent of $N$ at sufficiently large $N$. Finally, we say the dynamics is \emph{slow mixing} if $t_{\rm mix}\ge \exp[N^{c_{\rm slow}}]$ for a constant $c_{\rm slow}>0$ independent of $N$.

For any $\cL$ satisfying detailed balance \eqref{eq:detail_ba} (therefore preserves $\pi$), the weighted superoperator \begin{equation}\label{eq:H=piLpi}
    \cH[\cdot]:= -\pi^{-\frac{1}{4}} \cL[\pi^{\frac{1}{4}}\cdot \pi^{\frac{1}{4}}]\pi^{-\frac{1}{4}}
\end{equation}
is self-adjoint in the Hilbert–Schmidt inner product, so has real eigenvalues. This implies $-\cL$ has the same real eigenvalues because it is related by a similarity transformation. Since the channel $\ee^{t\cL}$ can only contract operators, the eigenvalues of $\cL$ are non-positive, and the largest one is $0$ from $\cL[\pi]=0$. Let $-\Delta$ be the second largest eigenvalue, then $\Delta^{-1}$ is the relaxation time, namely the asymptotic decay lifetimes of deviations around the thermal equilibrium. It upper bounds the mixing time by \cite{qmix_Temme13} 
\begin{equation}
    t_{\rm mix}\le \frac{1}{2\Delta}\lr{16+\log \norm{\pi^{-1}}}.
\end{equation}

As a result, for Gibbs state $\pi$, $\cL$ is fast mixing if the \emph{spectral gap} $\Delta$ is lower bounded by some polynomial of $N^{-1}$, because $\log \norm{\pi^{-1}}$ is polynomial in $N$ from \eqref{eq:Hu<1}. When $\Delta>0$ is a constant independent of $N$, it usually suggests (although does not prove) that $\cL$ is rapid mixing. 

\subsection{The classical case}

We say $H$ and its Gibbs state $\pi$ are classical, if there is a single-site basis $\{\ket{z_u}\}$ labeled by $z_u\in\{0,1,\cdots,q-1\}$ for each site $u\in \Lambda$ such that each Hamiltonian term $H_u$ is diagonal in that basis. The system then has a basis labeled by ``bitstring'' $\bz=\{z_u\}_{u\in \Lambda}$, and we use $\bz_S$ to denote the substring of $\bz$ on $S\subset \Lambda$. Operators
$H(\bz),\pi(\bz),\pi_S(\bz_S)$ etc are equivalently viewed as classical functions, and we say an operator algebra $\mathcal{A}$ is classical if it only contains operators diagonal in the $\bz$ basis.

The Markov property \eqref{eq:Ipi=0} in the classical case becomes the simple fact that $\forall S\subset \Lambda$, \begin{equation}
    \pi(\bz_S|\bz_{S^{\rm c}}) = \pi(\bz_S|\bz_{\partial S})
\end{equation}
where $\pi(\bz_S|\bz_{S^{\rm c}})$ for example is
the conditional probability of $\bz_S$ given $S^{\rm c}$. In other words, the condition distribution in a region is completely determined by the state on its neighbors. Sometimes we also write $\pi_S(\bz_S|\bz_{S^{\rm c}}):=\pi(\bz_S|\bz_{S^{\rm c}})$.

The Glauber dynamics generator $\cL$ \eqref{eq:glauber=} in the classical case becomes an operator instead of superoperator $\frac{\dd}{\dd t} \rho(\bz')= \sum_{\bz} \bra{\bz'}\cL\ket{\bz}\rho(\bz)$, where \begin{equation}\label{eq:Lu=}
    \cL_u = \sum_{z'_u,z_u,\bz_{\partial u}} \pi_u(z'_u|\bz_{\partial u}) \lr{\ket{z'_u,\bz_{\partial u}}\bra{z_u,\bz_{\partial u}}- \ket{z_u,\bz_{\partial u}}\bra{z_u,\bz_{\partial u}}}
\end{equation} 
resamples $u$ based on the state $\bz_{\partial u}$ of its neighbors. The detailed balance condition \eqref{eq:detail_ba} reduces to \begin{equation}\label{eq:DB}
    \bra{\bz'}\cL_u \ket{\bz}\pi(\bz) = \bra{\bz}\cL_u \ket{\bz'}\pi(\bz').
\end{equation}

\subsection{The Ising and Potts models}

We will use the classical Ising and Potts models as demonstrating examples when developing our general framework.
The $q$-state Potts model $H^{\rm Potts}=\sum_{u\in \Lambda}H_u^{\rm Potts}$ on the square lattice is defined by \begin{equation}\label{eq:Potts}
    H_u^{\rm Potts} = -\frac{\beta}{2}\sum_{v:\alr{u,v}} \mathbb{I}[z_u=z_v]
\end{equation} 
where each $z_u=0,1,\cdots,q-1$, $\mathbb{I}[z_u=z_v]$ is the indicator function that returns $1$ ($0$) if $z_u=z_v$ ($z_u\neq z_v$), and $\alr{u,v}$ means the unordered $u,v$ are neighbors in the square lattice (not in $G$, although all results would be qualitatively the same for that choice). There are $2d$ $v$s involved in the sum. The $q=2$ case reduces to the Ising model $H^{\rm Ising}$, where we also introduce an external field: \begin{equation}\label{eq:Ising}
    H_u^{\rm Ising} = \beta h z_u-\frac{\beta}{4}\sum_{v:\alr{u,v}} (1-2z_u)(1-2z_v)
\end{equation}
favoring $z_u=0$ for $h>0$.

\section{Strong decay of correlations and other mixing conditions}\label{sec:SDOC}

\subsection{Strong decay of correlations}

This paper studies consequences of the following condition.

\begin{defn}[Strong decay of correlations]\label{def:SDOC}
    Let $f=f(\cdot)$ be a monotonically non-increasing function and $R_{\rm A}\ge 1$. We say classical Gibbs state $\pi$ satisfies $(R_{\rm A},f)$-strong-decay-of-correlations (SDOC)
    , if for any region $A\subset \Lambda$ with bounded diameter \begin{equation}\label{eq:A<cA}
        \diam(A)\le R_{\rm A}
    \end{equation}
    and for any $r\ge 1$ with region $C=\{u\in \Lambda: \mathrm{dist}(u,A)\ge r\}$,
    we have
    \begin{equation}\label{eq:SDOC}
        \max_{\bz_C} \norm{\pi_A(\cdot|\bz_C)-\pi_A}_1 \le f(r)
    \end{equation}
    maximized over all bitstrings $\bz_C$ on $C$. Here  $\norm{\pi_A(\cdot|\bz_C)-\pi_A}_1 =\sum_{\bz_A} \abs{\pi_A(\bz_A|\bz_C)-\pi_A(\bz_A)}$.
\end{defn}

Intuitively, \eqref{eq:SDOC} says the conditional Gibbs measure in region $A$ is almost thermal as long as it is far from the conditioned region $C$.
We will work with the case where $R_{\rm A}$ is a constant independent of system size and function $f(r)$ decays sufficiently fast, e.g. as a sufficiently fast power function.

The roles of $A,C$ in quantity \eqref{eq:SDOC} can be symmetrized to yield quantity
\begin{equation}\label{eq:ratio_cor}
    \norm{\frac{\pi_{AC}}{\pi_A \otimes \pi_C} -1}= \max_{\bz_A,\bz_C} \abs{\frac{\pi_{AC}(\bz_A,\bz_C)}{\pi_A(\bz_A) \cdot \pi_C(\bz_C)} -1}=\max_{\bz_A,\bz_C} \abs{\frac{\pi_A(\bz_A|\bz_C)}{\pi_A(\bz_A)} -1}.
\end{equation}
The following Proposition shows that this ``ratio-correlation'' is of the same order as \eqref{eq:SDOC} for constant $R_{\rm A}$, so we could have also formulated SDOC using this symmetrized version.

\begin{prop}\label{prop:ratio_cor}
    In a classical Gibbs state with \eqref{eq:A<cdd}, the two quantities above can differ by at most a constant determined by $d,q,\beta,R_{\rm A}$: \begin{equation}\label{eq:vs_ratio}
    \lr{q\ee^{2\cdot 3^d\beta}}^{-c_d R_{\rm A}^d}\norm{\frac{\pi_{AC}}{\pi_A \otimes \pi_C} -1} \le \max_{\bz_C} \norm{\pi_A(\cdot|\bz_C)-\pi_A(\cdot)}_1\le \norm{\frac{\pi_{AC}}{\pi_A \otimes \pi_C} -1}.
\end{equation}
\end{prop} 

\begin{proof}
    
The second inequality in \eqref{eq:vs_ratio} is straightforward by comparing \eqref{eq:SDOC} and \eqref{eq:ratio_cor}. The first comes from \begin{equation}
    \abs{\frac{\pi_{AC}(\bz_A,\bz_C)}{\pi_A(\bz_A) \cdot \pi_C(\bz_C)} -1} \le \frac{1}{\pi_A(\bz_A)} \sum_{\bz'_A}\abs{\frac{\pi_{AC}(\bz'_A,\bz_C)}{\pi_C(\bz_C)} - \pi_A (\bz'_A) } 
\end{equation}
and \begin{align}\label{eq:piA>}
    \frac{1}{\pi_A(\bz_A)}&=\sum_{\bz'_A}\frac{\pi_A(\bz'_A)}{\pi_A(\bz_A)} \nonumber\\
    &=\sum_{\bz'_A}\frac{\sum_{\bz_{A^{\rm c}}}\ee^{-H(\bz'_A,\bz_{A^{\rm c}})}}{\sum_{\bz_{A^{\rm c}}}\ee^{-H(\bz_A,\bz_{A^{\rm c}})}}\nonumber\\
    &\le \sum_{\bz'_A}\max_{\bz_{A^{\rm c}}}\ee^{-H(\bz'_A,\bz_{A^{\rm c}})+H(\bz_A,\bz_{A^{\rm c}})} \nonumber\\
    &\le \sum_{\bz'_A}\ee^{2\sum_{u:H_u \text{ acts on }A} \norm{H_u} } \nonumber\\
    &\le q^{|A|}\cdot \ee^{2\cdot 3^d \beta|A|}\nonumber\\
    &\le \lr{q\ee^{2\cdot 3^d\beta}}^{c_d R_{\rm A}^d}
\end{align}
using \eqref{eq:Hu<1} and the fact that each vertex is acted on by at most $3^d$ $H_u$ terms. The last line of \eqref{eq:piA>} utilizes \begin{equation}\label{eq:A<cdd}
    |A|\le c_d R_{\rm A}^d
\end{equation} 
from \eqref{eq:cd} and \eqref{eq:A<cA}. 

\end{proof}

The quantity \eqref{eq:ratio_cor} being small without requiring a bounded $\diam(A)$ is referred to as a ``mixing condition'' in \cite{weak_mixing98,strong_decay_anyd24}. However, we call \eqref{eq:SDOC} SDOC in this work to emphasize the boundedness of $A$ (we will explain more below by comparing to a slightly different mixing condition). The term ``strong'' in Definition \ref{def:SDOC} comes from the fact that
it is stronger than the usual decay of point-to-set connected correlation functions or mutual information, as shown by the next subsection.

\subsection{Comparing to other point-to-set correlation measures}

\begin{prop}[Adapted from Section 3.1 in \cite{strong_decay_1d22}]\label{prop:cor_measures}
    For any full-rank quantum state $\pi_{AC}$ (not necessarily part of a Gibbs state) on two subsystems $A,C$ with reduced density matrices $\pi_A,\pi_C$ (their inverses being $\pi_A^{-1},\pi_C^{-1}$), \begin{equation}\label{eq:cor<I<}
        \frac{1}{2}\mathrm{Corr}_\pi(A,C)^2 \le \frac{1}{2} \norm{\pi_{AC}-\pi_A\otimes \pi_C}_1^2 \le I_\pi(A:C) \le \norm{\pi_A^{-1}\pi_C^{-1}\pi_{AC}-1} .
    \end{equation}
    Here \begin{equation}
        \mathrm{Corr}_\pi(A,C):= \mathop{\rm sup}\limits_{\OO_A,\OO'_C} \abs{\alr{\OO_A\OO'_C}_\pi-\alr{\OO_A}_\pi\alr{\OO'_C}_\pi}
    \end{equation}
    where $\OO_A,\OO'_C$ are operators supported on $A,C$ respectively with $\norm{\OO_A}=\norm{\OO'_C}=1$, and \begin{equation}\label{eq:IAC=}
        I_\pi(A:C) = \alr{\log\pi_{AC}-\log(\pi_A\otimes \pi_C)}_\pi
    \end{equation}
    is the quantum mutual information. If $\pi$ is a classical distribution, $I_\pi(A:C)$ becomes the classical mutual information, and the last expression in \eqref{eq:cor<I<} becomes \eqref{eq:ratio_cor}. 
\end{prop}

The last two inequalities in \eqref{eq:cor<I<} can be qualitatively loose at large $|C|$ even for classical states and $|A|=\order(1)$. For the last inequality, here is an example: 

\begin{eg}
    Suppose $A$ is a single bit with values $0,1$ and $C$ takes values $0,1,\cdots,n$, so that the states are of the form $(z_A,z_C)$. Let $\pi'_{AC}$ be the uniform classical distribution on all states, and full-rank $\pi_{AC}=\frac{1}{n}\delta_{(0,0)}+(1-\frac{1}{n})\pi'_{AC}$ where $\delta_{(0,0)}$ is the distribution supported on only $(0,0)$. Then if $n$ is larger than some constant, \begin{equation}
        I_\pi(A:C) \le \frac{1}{n}
    \end{equation}
    while \begin{equation}
        \norm{\frac{\pi_{AC}}{\pi_A \otimes \pi_C} -1} \ge \frac{1}{3}.
    \end{equation}
\end{eg}

The idea here is that rare events $z_C$ does not contribute much to mutual information, yet conditioning on the rare event could significantly influence $A$. A Gibbs state example of this fact is the 2d Ising model with a weak field $h=N^{-1/8}$: The nonzero field makes the global equilibrium $\pi$ to be mostly supported in the trivial phase with positive polarization $1-2z_u$, so mutual information is expected to decay exponentially with distance. However, conditioned on rare events where $z_u\equiv 1$ in $C$, $A$ will be flipped to negative polarization as long as $|C^{\rm c}|\ll N^{1/8}$, destroying SDOC.

\begin{figure}
    \centering
    \includegraphics[width=0.9\linewidth]{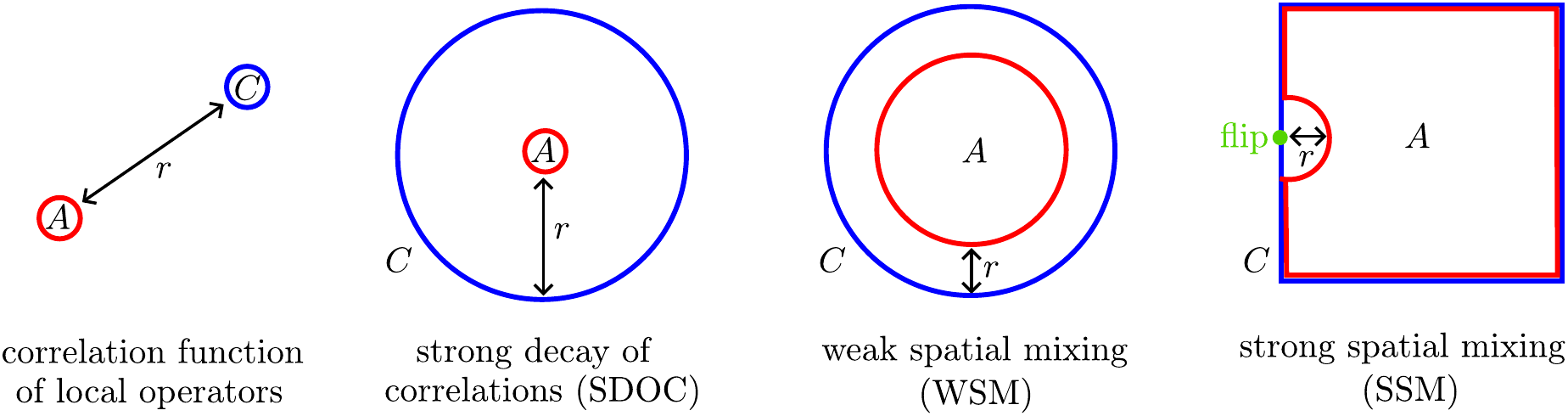}
    \caption{Geometry for different correlation decay properties in classical Gibbs states, which become stronger when going from left to right. In the first three cases, one asks for the influence in region $A$ when conditioning $C$ in any bitstring $\bz_C$. For the last SSM, one compares the influence of flipping a boundary bit (green) in any $\bz_C$. }
    \label{fig:cors}
\end{figure}

Note that for point-to-point correlation with $|A|=|C|=\order(1)$, the different measures in \eqref{eq:cor<I<} (where the last $\norm{\pi_A^{-1}\pi_C^{-1}\pi_{AC}-1}$ is replaced by its square) can differ by at most a constant prefactor depending on $d,\beta$, as one can verify using lower bounds on marginal probabilities like \eqref{eq:piA>}. So to characterize point-to-point correlation in Gibbs states, connected correlation functions are sufficient. See Fig.~\ref{fig:cors} for an illustration of some of the different correlation measures including the ones to be introduced. 

\subsection{Weak spatial mixing}
SDOC captures point-to-set correlations where the outer set $C$ is conditioned on. Its set-to-set version is the weak spatial mixing condition \cite{Martinelli99} \footnote{There it is called weak mixing. However, to separate from temporal mixing of dynamics, we follow \cite{SSM_rev04} to add the word ``spatial''.}.

\begin{defn}[Weak spatial mixing]\label{def:WSM}
    We say classical Gibbs state $\pi$ satisfies $f_{\rm WSM}$-weak-spatial-mixing (WSM), if for any subsets $A,C\subset \Lambda$ with $r=\mathrm{dist}(A,C)\ge 1$, the Gibbs measure on $A$ conditioned on a fixed bitstring in $C$ obeys \begin{equation}\label{eq:WSM}
        \max_{\bz_C,\bz'_C}\norm{\pi_A(\,\cdot\,|\bz_C) - \pi_A(\,\cdot\,|\bz'_C)  }_1 \le |A||\partial(C^{\rm c})| f_{\rm WSM}(r)
    \end{equation}
    where $\partial(C^{\rm c})=\{u\in C: \mathrm{dist}(u,C^{\rm c})=1\}$ is the inner-boundary of $C$.
\end{defn}

In Definition 2.3 in \cite{Martinelli99}, the right hand side of \eqref{eq:WSM} is of a slightly different form $c_1 \sum_{u\in A,v\in \partial(C^{\rm c})}\ee^{-c_2\mathrm{dist}(u,v)}$ with constants $c_1,c_2$, which can be upper bounded by the right hand side in \eqref{eq:WSM} with $f(r)=c_1 \ee^{-c_2 r}$. We call such a case exponential WSM. WSM \eqref{eq:WSM} implies \begin{equation}\label{eq:WSM'}
    \max_{\bz_C}\norm{\pi_A(\,\cdot\,|\bz_C) - \pi_A  }_1 \le |A||\partial(C^{\rm c})| f_{\rm WSM}(r),
\end{equation}
i.e. $(R_{\rm A},f)$-SDOC for any $R_{\rm A}$ with \begin{equation}\label{eq:fr=WSM}
    f(r)=c_d^3 R_{\rm A}^{2d}  \cdot  (r+1)^{d-1} f_{\rm WSM}(r)
\end{equation}
using \eqref{eq:A<cdd}, \eqref{eq:cd} and $|\partial(C^{\rm c})|\le |A|\times$ (number of vertices of distance exactly $r$ to a given vertex). 

On the other hand, we do not know of a Gibbs state example that satisfies SDOC but not WSM with a comparable decay function. Nevertheless, we choose the potentially weaker SDOC condition for our latter results because the physical picture is clearly point-to-set like, as we will see.

\subsection{Strong spatial mixing and boundary order}\label{sec:SSM}

Perhaps the strongest notion of correlation decay in Gibbs states is the strong spatial mixing (SSM) condition \cite{Martinelli99,SSM_rev04}, which we overview in this subsection in order to compare our results with prior work. Here we define SSM following \cite{plaquet17}:

\begin{defn}[Strong spatial mixing]
    We say classical Gibbs state $\pi$ satisfies $(c'_{\rm SSM},c_{\rm SSM})$-strong-spatial-mixing (SSM) with constants $c'_{\rm SSM},c_{\rm SSM}$ if the following holds. Consider any $L\times L\times \cdots\times L$ box $S\subset\Lambda$ on the square lattice and any neighboring site $u\in \partial S$ and $r\ge 1$. Let $A:=\{v\in S:\mathrm{dist}(u,v)\ge r\}$. For any bitstring pair $\bz_{S^{\rm c}},\bz'_{S^{\rm c}}$ outside the box that only differ at site $u$, \begin{equation}\label{eq:SSM}
        \norm{\pi_A(\,\cdot\,|\bz_{S^{\rm c}}) - \pi_A(\,\cdot\,|\bz'_{S^{\rm c}})  }_1 \le c'_{\rm SSM}\ee^{-c_{\rm SSM}\cdot r}.
    \end{equation}
\end{defn}

SSM is stronger than WSM in $d\ge 3$ \cite{WSMSSM_2d94,Martinelli99}, and therefore stronger than all point-to-set correlation conditions in Proposition \ref{prop:cor_measures}. Intuitively, for a system obeying SSM, the difference at boundary condition $u$ cannot influence region $A$ faraway from $u$, even if $A$ is allowed to touch the conditioned region $S^{\rm c}$ (which is not the case for WSM). In other words, a system obeying WSM may violate SSM from ``boundary order'', which is believed to be the case for the 3d Ising model \eqref{eq:Ising} at low temperature and certain values of the magnetic field \cite{Martinelli99,Ising_3dboundary07}, and has a provable example \cite{boundary_order_Czech86}. The physical picture for the 3d Ising example is as follows. Consider one face $F$ of the $L\times L\times \cdots \times L$ box to have $z_u=1$ boundary conditions, while the other five faces are free (i.e. turning off the interactions across those faces). Let the field $h>0$ prefers $z_u=0$, the conditional Gibbs state inside at low temperature is mostly zero with a thin layer of $1$s attached to $F$ with width $w_h$ determined by $h$. Moreover, the surface separating the $0$s and $1$s almost does not fluctuate. One can then fine tune $h$ so that $w_h$ and $w_h+1$ becomes equally likely, forming the boundary order.

As shown by e.g. \cite{SSM_rev04}, SSM implies rapid mixing of Glauber dynamics, while the reversed direction requires rapid mixing for every boundary condition:

\begin{thm}\label{thm:SSM}
    If classical Gibbs state $\pi$ satisfies $(c'_{\rm SSM},c_{\rm SSM})$-SSM, then its Glauber dynamics \eqref{eq:L=Lu} with \eqref{eq:Lu=} mixes rapidly, where $c_{\rm rapid}$ in \eqref{eq:rapid} is determined by $d,q,\beta,c'_{\rm SSM},c_{\rm SSM}$. 
    
    On the other hand, if there exists a constant $c_{\rm rapid}$ such that for any subset $S\subset \Lambda$ and any bitstring $\bz_{S^{\rm c}}$ on its complement, the Glauber dynamics $\sum_{u\in S}\cL_u$ in $S$ with $\bz_{S^{\rm c}}$ fixed as boundary condition mixes rapidly to $\pi_S(\cdot|\bz_{S^{\rm c}})$ with mixing time $\le c_{\rm rapid}\log|S|$, then the classical Gibbs state $\pi$ satisfies $(c'_{\rm SSM},c_{\rm SSM})$-SSM with constants $c'_{\rm SSM},c_{\rm SSM}$ determined by $d,q,\beta,c_{\rm rapid}$. 
\end{thm}

For the second part of this Theorem, rapid mixing for the global $\cL$ alone is not sufficient for SSM, as suggested by the 3d Ising model above: 
As a monotonic model, it is rapid mixing at any $h\neq 0$ due to WSM \cite{Ising_WSM_nonzeroh19} for e.g. periodic boundary conditions, yet is expected to not satisfy SSM.
This example stresses the discrepancy between SSM and temporal mixing:
Although the presence of boundary order may cause the Glauber dynamics to mix slowly if the boundary sites are kept fixed, it does not exclude the possibility that Glauber dynamics under other boundary conditions like free or periodic boundary conditions mix rapidly. The latter boundary conditions are more natural in the sense that they do not deliberately keep sites fixed; more generally, it is plausible that the ``bulk'' of such models far from fixed boundaries mixes rapidly independent of the boundary condition. Indeed, to the best of our knowledge there is no example that satisfy WSM (or even SDOC) with a fast decay function, and yet does not mix rapidly in the bulk. In this paper, we make progress on showing this general relation, which relaxes the SSM-type conditions required in prior works showing rapid/fast mixing for general non-monotonic systems.

\section{Strong decay of correlations implies quasi-local connection}\label{sec:connection}

\subsection{Coherent Gibbs state}

For the classical case, there is a simpler quantum pure state representation of the Gibbs state comparing to the canonical purification \eqref{eq:purification}. The classical distribution $\pi$ can be written in a normalized quantum state \cite{RokhsarKivelson88,RKHamiltonian_generalize05} \begin{equation}\label{eq:kpi=}
    \kpi = \sum_{\bz} \sqrt{\pi(\bz)} \ket{\bz} = \mathcal{Z}^{-\frac{1}{2}}\sum_{\bz} \ee^{-\frac{1}{2}H(\bz)} \ket{\bz} =: \mathcal{Z}^{-\frac{1}{2}} \kpi_{\rm u}
\end{equation}
with non-negative amplitudes, which we call the coherent Gibbs state. Here $\mathcal{Z}=\trace\lr{\ee^{-H}}$ is the partition function, and $\kpi_{\rm u}$ is the unnormalized version. For example, the infinite temperature state proportional to the identity operator is coherently written as \begin{equation}
    \kplus\propto \sum_\bz \ket{\bz}
\end{equation}
normalized by $\braket{\bm{+}|\bm{+}}=1$. It is a product state $\ket{\bm{+}}=\otimes_u \ket{+}_u$. We will call this state the all-$+$ state suggested by the case of qubits. 

The coherent Gibbs state is related to the canonical purification \eqref{eq:purification} by \begin{equation}\label{eq:kpi=kpi}
    \kett{\sqrt{\pi}} = \otimes_{u\in \Lambda} \mathrm{CNOT}_u \lr{\kpi\otimes \ket{\bm{0}} }
\end{equation}
where $\mathrm{CNOT}_u$ is the controlled unitary acting on $u$ and its ancilla that rotates the ancilla from $0$ to $z_u$ if the original copy is $z_u$. As a result, if two $\kpi$s are connected by a quasi-local unitary so do the two $\kett{\sqrt{\pi}}$s. Intuitively, $\kett{\sqrt{\pi}}$ is just $\kpi$ cloned (locally) to two copies.

\subsection{Main result}

\begin{thm}[Strong decay of correlations implies quasi-local connection]\label{thm:cohe}
    Consider a path of classical Hamiltonians $H(s)=\sum_{u\in \Lambda} H_{u}(s)$ ($s\in[0,1]$) where each $H_{u}(s)$ is differentiable with \begin{equation}\label{eq:dsH<1}
        \norm{\partial_s H_u(s)} \le \beta'.
    \end{equation}
    If at any $s$ in the path, \eqref{eq:Hu<1} holds for $H=H(s)$ and the Gibbs state $\pi_s$ at $s$ satisfies $(2,f)$-SDOC where function $f$ does not depend on $s$. Then for any integer $\ell_{\rm max}>1$, the following two statements hold: \begin{enumerate}
        \item The coherent Gibbs state version of \eqref{eq:pi=Upi} holds:
    \begin{equation}\label{eq:kpi=Ukpi}
        \ket{\sqrt{\pi_{1}}} = \mathcal{T} \ee^{-\ii \int^{1}_{0} W(s) \dd s} \ket{\sqrt{\pi_0}}
    \end{equation}
    where \begin{equation}\label{eq:W=suml}
        W(s)=\sum_{u\in \Lambda}\sum_{\ell=1}^{\ell_{\rm max}} W_{u,\ell}(s)
    \end{equation} 
    is Hermitian. $W_{u,\ell}(s)$ ($\ell<\ell_{\rm max}$) acts no further than distance $2^\ell$ to $u\in \Lambda$, and decays with $\ell$: \begin{equation}\label{eq:W_decay}
        \norm{W_{u,1}(s)}\le \beta',\qquad \norm{W_{u,\ell}(s)} \le \beta'\cdot f(2^{\ell-1}-1) \quad (\ell>1).
    \end{equation}
    \eqref{eq:W_decay} holds also for $\ell=\ell_{\rm max}$.

    \item The states are connected in both directions by quasi-local Lindbladians in \eqref{eq:2way} with \begin{equation}\label{eq:L=suml}
        \cL^{1\leftarrow 0}(s)=\sum_{u\in \Lambda}\sum_{\ell=1}^{\ell_{\rm max}} \cL^{1\leftarrow 0}_{u,\ell}(s)
    \end{equation}
    where $\cL^{1\leftarrow 0}_{u,\ell}(s)$ acts in the same range as $W_{u,\ell}(s)$ above, and \begin{equation}\label{eq:L_decay}
        \norm{\cL^{1\leftarrow 0}_{u,1}(s)}_\diamond\le 4\beta',\qquad \norm{\cL^{1\leftarrow 0}_{u,\ell}(s)}_\diamond \le 4\beta'\cdot f(2^{\ell-1}-1) \quad (\ell>1)
    \end{equation}
    for diamond norm $\norm{\cL}_\diamond:=\max_{\OO:\norm{\OO}_1=1} \norm{(\cL\otimes I)[\OO]}_1$.
    The same locality bounds hold for $\cL^{0\leftarrow 1}$ as well.
    \end{enumerate}
\end{thm}

We introduce $\ell_{\rm max}$ so that if function $f(r)$ ceases decaying at some large $r$, we can set $2^{\ell_{\rm max}}$ to be of the order of that value. For example, if $2^{\ell_{\rm max}}\gg \log N$ and $f(r)$ decays exponentially before that value, the nonlocal $W_{u,\ell_{\rm max}}$ would be polynomially small in $N$ which would not destroy the quasi-locality much. We restrict terms in $W$ to be supported only in radii of powers of two, so that a tighter Lieb-Robinson bound can be applied for its evolution.

\subsection{Proof of quasi-local connection}

\begin{proof}[Proof of Theorem \ref{thm:cohe}]
We first prove the coherent Gibbs state conclusion.
We consider $H,\kpi$ as an implicit function of $s$. Taking derivative over $s$, \eqref{eq:kpi=} implies \begin{align}
    \partial_s \kpi_{\rm u} &= -\frac{1}{2} \sum_\bz \partial_sH(\bz) \ee^{-\frac{1}{2}H(\bz)} \ket{\bz} \nonumber\\
    &= -\frac{1}{2}\lr{\partial_s H} \kpi_{\rm u}
\end{align}
where $\partial_s H$ in the last line is the corresponding quantum operator. The normalized state then changes by \begin{align}\label{eq:dpi=}
    \partial_s \kpi &= (I-\kpi \bra{\sqrt{\pi}})\mathcal{Z}^{-1/2} \partial_s \kpi_{\rm u} \nonumber\\
    &= -\frac{1}{2}\lr{\partial_sH-\alr{\partial_s H}_\pi} \kpi \nonumber\\
    &= -\frac{1}{2}\sum_{u\in \Lambda} \lr{\partial_sH_u-\alr{\partial_sH_u}_\pi} \kpi
\end{align}
where for example $\alr{\partial_sH}_\pi = \bra{\sqrt{\pi}}\partial_s H \kpi$.

\subsubsection{Separating terms using strong decay of correlations}

We first focus on one $u$ and aim to show \begin{equation}\label{eq:=sumVr}
    \partial_sH_u-\alr{\partial_sH_u}_\pi = \sum_{\ell=1}^{\ell_{\rm max}} V_{\ell}
\end{equation}
with Hermitian $V_{\ell}$s (ignoring the $u$ subscript) that satisfy good properties: (\emph{1}) they are quasi-local and (\emph{2}) they satisfy certain orthogonality properties when acting on $\kpi$. 

We first explain the quasi-locality.
$V_\ell$ ($\ell\le \ell_{\rm max}-1$) is 
supported in the radius-$L$ ball \begin{equation}
    B_L:=\{v: \mathrm{dist}(u,v)\le L\}
\end{equation} 
together with its neighborhood $\partial B_L:=\partial \lr{B_L}$ (i.e. supported in the radius-$(L+1)$ ball), where \begin{equation}\label{eq:L=2l}
    L:=2^\ell-1.
\end{equation} 
$V_{\ell_{\rm max}}$ is supported in the whole system. To define those $V_\ell$, we first for each $\ell$ consider the expectation of $\partial_s H_u$ \begin{equation}\label{eq:tV=}
    \widetilde{V}_{\ell}(\bz_{\partial B_L}) = \pi\lr{\partial_sH_u|\bz_{\partial B_L} }.
\end{equation}
conditioned on $\bz_{\partial B_L}$. 
We apply SDOC \eqref{eq:SDOC} for $A=$ support of $H_u$ (since 
$\diam(A)\le2$ is in the assumed SDOC range), $C=B_L^{\rm c}$ with 
$\mathrm{dist}(A,C)\ge L$ to get \begin{align}\label{eq:<2f}
    \abs{\widetilde{V}_{\ell}(\bz_{\partial B_L}) - \alr{\partial_sH_u}_\pi} &\le \norm{\partial_s H_u}\cdot \norm{\pi_A(\cdot|\bz_C)-\pi_A}_1 \le \beta'\cdot f(L)
\end{align}
using \eqref{eq:dsH<1}.
Since $\widetilde{V}_{\ell}$ can be interpreted as a diagonal operator acting on $\partial B_L$, this implies \begin{equation}\label{eq:V-s<}
    \norm{\widetilde{V}_{\ell} - \alr{\partial_sH_u}_\pi} \le \beta'\cdot f(L)=\beta'\cdot f(2^\ell-1).
\end{equation}
We then define \begin{align}\label{eq:V=tV}
    V_{1} &= \partial_sH_u-\widetilde{V}_{1}, \nonumber\\
    V_{2} &= \widetilde{V}_{1}-\widetilde{V}_{2}, \nonumber\\
    V_{3} &= \widetilde{V}_{2}-\widetilde{V}_{3}, \cdots, \nonumber\\
    V_{\ell_{\rm max}} &= \widetilde{V}_{\ell_{\rm max}-1}-\alr{\partial_sH_u}_\pi .
\end{align}
\eqref{eq:=sumVr} then holds because \begin{equation}
    \partial_sH_u-\alr{\partial_sH_u}_\pi = (\partial_sH_u-\widetilde{V}_{1}) + (\widetilde{V}_{1}-\widetilde{V}_{2})+ \cdots + (\widetilde{V}_{\ell_{\rm max}-1}-\alr{\partial_sH_u}_\pi ).
\end{equation}
Furthermore, \begin{equation}\label{eq:VLmin}
    \norm{V_{1}} \le \beta'+ \norm{\widetilde{V}_{1}} \le 2\beta'
\end{equation}
using \eqref{eq:dsH<1},
while for all $\ell > 1$ including $\ell_{\rm max}$, \begin{equation}\label{eq:Vr<}
    \norm{V_\ell} \le \norm{\widetilde{V}_{\ell} - \alr{\partial_sH_u}_\pi} + \norm{\widetilde{V}_{\ell-1} - \alr{\partial_sH_u}_\pi} \le 2 \beta'\cdot f(2^{\ell-1}-1)
\end{equation}
from triangle inequality, \eqref{eq:V-s<} and monotonicity of $f(L)$. 

\subsubsection{Orthogonality properties}

We then demonstrate the orthogonality properties: From Markov property of Gibbs state, \begin{equation}\label{eq:pi=sum_psi}
    \kpi = \sum_{\bz_{B^{\rm c}_L}} \sqrt{\pi(\bz_{B^{\rm c}_L})} \ket{\bz_{B^{\rm c}_L}} \otimes \sum_{\bz_{B_L}} \sqrt{\pi(\bz_{B_L}|\bz_{\partial B_L})} \ket{\bz_{B_L}} =: \sum_{\bz_{B^{\rm c}_L}} \sqrt{\pi(\bz_{B^{\rm c}_L})} \ket{\bz_{B^{\rm c}_L}} \otimes \ket{\psi(\bz_{\partial B_L})}
\end{equation}
(note that $\ket{\psi(\bz_{\partial B_L})}$ is a state on $B_L$, not $\partial B_L$)
while \begin{equation}\label{eq:Vpi=phi}
    V_\ell\kpi = \sum_{\bz_{B^{\rm c}_L}} \sqrt{\pi(\bz_{B^{\rm c}_L})} \ket{\bz_{B^{\rm c}_L}} \otimes \sum_{\bz_{B_L}} V_\ell(\bz_{\overline{B_L}})\sqrt{\pi(\bz_{B_L}|\bz_{\partial B_L})} \ket{\bz_{B_L}}=: \sum_{\bz_{B^{\rm c}_L}} \sqrt{\pi(\bz_{B^{\rm c}_L})} \ket{\bz_{B^{\rm c}_L}} \otimes \ket{\phi(\bz_{\partial B_L})}
\end{equation}
($\overline{B_L}=B_L\sqcup \partial B_L$) so that \begin{align}
    \braket{\psi(\bz_{\partial B_L})| \phi(\bz_{\partial B_L})} &= \sum_{\bz_{B_L}} V_\ell(\bz_{\overline{B_L}}) \pi(\bz_{B_L}|\bz_{\partial B_L}) \nonumber\\
    &= \sum_{\bz_{B_L}} \mlr{\pi\lr{\partial_sH_u|\bz_{\partial B_{L'}} }-\pi\lr{\partial_sH_u|\bz_{\partial B_L} } }\pi(\bz_{B_L}|\bz_{\partial B_L})  \hspace{10em}\mytag{\text{Using \eqref{eq:V=tV} and \eqref{eq:tV=}}} \\
    &= \lr{\sum_{\bz_{\partial B_{L'}}}\pi\lr{\partial_sH_u|\bz_{\partial B_{L'}} }\pi(\bz_{\partial B_{L'}}|\bz_{\partial B_L})}-\pi\lr{\partial_sH_u|\bz_{\partial B_L} } \nonumber\\
    &=0\label{eq:psiphi=0}
\end{align}
for $1<\ell<\ell_{\rm max}$, where $L'=2^{\ell-1}-1$.
Here the last line comes from the fact that sampling $\partial_sH_u$ conditioned on $\bz_{\partial B_L}$, is equivalent to first sampling $\bz_{\partial B_{L'}}$ conditioned on $\bz_{\partial B_L}$, then sampling $\partial_sH_u$ conditioned on $\bz_{\partial B_{L'}}$ due to Markov property. One can verify that \eqref{eq:psiphi=0} also holds for $\ell=1$.
This orthogonality condition ensures that we can define Hermitian \begin{equation}\label{eq:W=V}
    W_\ell = -\frac{\ii}{2} \sum_{\bz_{\partial B_L}}\ket{\bz_{\partial B_L}}\bra{\bz_{\partial B_L}}\otimes \lr{\ket{\phi(\bz_{\partial B_L})}\bra{\psi(\bz_{\partial B_L})} - \ket{\psi(\bz_{\partial B_L})}\bra{\phi(\bz_{\partial B_L})}}
\end{equation}
supported in $\overline{B_L}$ (same as $V_\ell$) for all $\ell<\ell_{\rm max}$,
such that \begin{align}
    W_\ell \kpi &= W_\ell \sum_{\bz_{B^{\rm c}_L}} \sqrt{\pi(\bz_{B^{\rm c}_L})} \ket{\bz_{B^{\rm c}_L}} \otimes \ket{\psi(\bz_{\partial B_L})} \nonumber\\
    &=-\frac{\ii}{2} \sum_{\bz_{B^{\rm c}_L}} \sqrt{\pi(\bz_{B^{\rm c}_L})} \ket{\bz_{B^{\rm c}_L}} \otimes \lr{\ket{\phi(\bz_{\partial B_L})}\bra{\psi(\bz_{\partial B_L})} - \ket{\psi(\bz_{\partial B_L})}\bra{\phi(\bz_{\partial B_L})}}\ket{\psi(\bz_{\partial B_L})} \nonumber\\
    &= -\frac{\ii}{2} \sum_{\bz_{B^{\rm c}_L}} \sqrt{\pi(\bz_{B^{\rm c}_L})} \ket{\bz_{B^{\rm c}_L}} \otimes \ket{\phi(\bz_{\partial B_L})} \mytag{\text{Using \eqref{eq:psiphi=0}}}\\
    &= -\frac{\ii}{2} V_\ell \kpi. \label{eq:WL=iVL}
\end{align}
On the other hand, \eqref{eq:W=V} implies
\begin{align}
    \norm{W_\ell} &\le \frac{1}{2} \max_{\bz_{\partial B_L}} \norm{\ket{\phi(\bz_{\partial B_L})} } \mytag{\psi(\bz_{\partial B_L})\text{ is normalized in \eqref{eq:pi=sum_psi}}} \\
    &= \frac{1}{2} \max_{\bz_{\partial B_L}} \sqrt{\sum_{\bz_{B_L}} \lr{V_\ell(\bz_{\overline{B_L}})}^2 \pi(\bz_{B_L}|\bz_{\partial B_L})} \mytag{\text{Using \eqref{eq:Vpi=phi}}} \\
    &\le \frac{1}{2} \max_{\bz_{\partial B_L}} \max_{\bz_{B_L}} \abs{V_\ell(\bz_{\overline{B_L}})} \mytag{\sum_{\bz_{B_L}} \pi(\bz_{B_L}|\bz_{\partial B_L})=1} \\
    &= \frac{1}{2} \norm{V_\ell}.\label{eq:Wr<}
\end{align}
For the last $\ell=\ell_{\rm max}$, we do not care about its locality so just define \begin{equation}
    W_{\ell_{\rm max}} = -\frac{\ii}{2} \mlr{V_{\ell_{\rm max}}, \kpi\bra{\sqrt{\pi}}}
\end{equation}
that also satisfies \eqref{eq:WL=iVL} and \eqref{eq:Wr<}, as one can verify using $\bra{\sqrt{\pi}}V_{\ell_{\rm max}}\kpi = 0$.

\eqref{eq:=sumVr} then becomes \begin{equation}
    \lr{\partial_sH_u-\alr{\partial_sH_u}_\pi} \kpi = 2\ii \sum_{\ell=1}^{\ell_{\rm max}} W_{u,\ell} \kpi,
\end{equation}
where we have put the $u$ subscript back $W_\ell \rightarrow W_{u,\ell}$. \eqref{eq:dpi=} then becomes \begin{align}
    \partial_s \kpi &= -\ii \sum_{u\in \Lambda}\sum_{\ell= 1}^{\ell_{\rm max}} W_{u,\ell} \kpi =: -\ii W \kpi,
\end{align}
which integrates to \eqref{eq:kpi=Ukpi}. The quasi-locality bound \eqref{eq:W_decay} of $W$ follows from combining \eqref{eq:Wr<}, \eqref{eq:VLmin} and \eqref{eq:Vr<}.

\subsubsection{Lindbladian connection}
Finally, we prove the second statement of Theorem \ref{thm:cohe}. Similar to \eqref{eq:dpi=}, we have \begin{equation}\label{eq:dspi=}
    \partial_s \pi=-\sum_{u\in \Lambda} \lr{\partial_sH_u-\alr{\partial_sH_u}_\pi} \pi.
\end{equation}
As above, we focus on one $u$ and \eqref{eq:=sumVr} becomes \begin{equation}
    -\lr{\partial_sH_u-\alr{\partial_sH_u}_\pi} \pi = -\sum_{\ell=1}^{\ell_{\rm max}} V_\ell\cdot \pi 
\end{equation}
where $V_\ell$ is the same classical function as before. For each given $\bz_{\partial B_L}$, \eqref{eq:Vpi=phi} becomes \begin{align}
    -\lr{V_\ell\cdot \pi}(\bz_{B_L}|\bz_{\partial B_L}) &= -V_\ell(\bz_{\overline{B_L}})\cdot \pi(\bz_{B_L}|\bz_{\partial B_L}) \nonumber\\
    &=\norm{V_\ell} \mlr{ \sigma_\ell(\bz_{B_L}|\bz_{\partial B_L}) - \pi(\bz_{B_L}|\bz_{\partial B_L}) } \label{eq:jump}
\end{align}
where 
\begin{equation}
    \sigma_\ell(\bz_{B_L}|\bz_{\partial B_L}):= \pi(\bz_{B_L}|\bz_{\partial B_L})\lr{1-\frac{V_\ell(\bz_{\overline{B_L}})}{\norm{V_\ell}}}
\end{equation}
($\sigma_\ell(\bz_{B_L}|\bz_{\partial B_L}):=\pi(\bz_{B_L}|\bz_{\partial B_L})$ for the special case $V_\ell=0$)
is a valid probability distribution of $\bz_{B_L}$. In particular, $\sigma_\ell(\bz_{B_L}|\bz_{\partial B_L})\ge 0$ from $V_\ell(\bz_{\overline{B_L}})\le \norm{V_\ell}$, and 
\begin{equation}
    \sum_{\bz_{B_L}} \sigma_\ell(\bz_{B_L}|\bz_{\partial B_L}) = 1- \frac{1}{\norm{V_\ell}} \pi\lr{V_\ell|\bz_{\partial B_L}} =1 
\end{equation}
from the orthogonality condition \eqref{eq:psiphi=0}. From \eqref{eq:jump}, $-V_\ell\cdot \pi=\cL^{1\leftarrow 0}_{u,\ell}[\pi]$ for a Lindbladian acting on $\overline{B_L}$ that resamples $\bz_{B_L}$ in $\sigma_\ell$ conditioned on $\bz_{\partial B_L}$, with strength \begin{equation}
    \norm{\cL^{1\leftarrow 0}_{u,\ell}}_\diamond \le 2 \norm{V_\ell}.
\end{equation}
Note that for $\ell=\ell_{\rm max}$ the above equations still hold without conditioning, i.e. $-V_{\ell_{\rm max}}\cdot \pi = \norm{V_{\ell_{\rm max}}}(\sigma_{\ell_{\rm max}}-\pi)$ just corresponds to resampling a global state $\sigma_{\ell_{\rm max}}(\bz)$.

Plugging in the previous $V_\ell$ bounds and putting all $u$ and the $s$ dependence together then proves the Lindbladian statement in Theorem \ref{thm:cohe}, where the reversed direction $0\leftarrow 1$ also follows. 

\end{proof}

Although we focus on finite spatial dimensions in this paper, which is crucial for the latter proof of fast mixing, we remark that the proof above directly generalizes to any interaction graph where SDOC is decaying with the graph distance. 

\section{Fast mixing}\label{sec:fast}

\subsection{Glauber dynamics and its Hamiltonian}

The Glauber dynamics \eqref{eq:Lu=} satisfies detailed balance in \eqref{eq:DB},
which makes \begin{equation}\label{eq:H=L}
    \bra{\bz'}\cH_u \ket{\bz}:= -\frac{1}{\sqrt{\pi(\bz')}} \bra{\bz'}\cL_u \ket{\bz} \sqrt{\pi(\bz)}
\end{equation}
real symmetric. $\cH=\sum_{u\in \Lambda} \cH_u$ is then a Hamiltonian with \begin{equation}
    \cH_u=\sum_{\bz_{\partial u}} \ket{\bz_{\partial u}}\bra{\bz_{\partial u}}\otimes \lr{I-\ket{\sqrt{\pi_u(\cdot|\bz_{\partial u})}}\bra{\sqrt{\pi_u(\cdot|\bz_{\partial u})}}}\ge 0
\end{equation}
acting on $u\sqcup \partial u$, and $\kpi$ being its unique frustration-free ground state with energy $0$: \begin{equation}
    \bra{\bz'}\cH_u \kpi = \sum_\bz -\frac{1}{\sqrt{\pi(\bz')}} \bra{\bz'}\cL_u \ket{\bz} \sqrt{\pi(\bz)}\cdot  \sqrt{\pi(\bz)}= -\frac{1}{\sqrt{\pi(\bz')}}\sum_\bz  \bra{\bz}\cL_u \ket{\bz'} \pi(\bz') = 0
\end{equation}
using \eqref{eq:H=L}, \eqref{eq:DB} and $\sum_\bz \bra{\bz}\cL_u=0$ (total probability conservation).

\subsection{Fast mixing results}

\begin{thm}\label{thm:gap}
    Consider coherent Gibbs state $\kpi=U\kplus$ where $\kplus$ is the all-$+$ state, $U$ satisfies the following Lieb-Robinson bound: For any Hermitian operator $\OO$ with $\norm{\OO}=1$ supported in $S\subset \Lambda$ and any $r\in \mathbb{Z}_{>0}$, there exist Hermitian operators $\OO'_r,\OO''_r$ supported within distance $r$ to $S$ such that it approximates the forward and backward Heisenberg evolution \begin{equation}\label{eq:LRB}
        \norm{U^\dagger \OO U - \OO'_r} \le |S| f_{\rm LR}(r), \qquad \norm{U\OO U^\dagger - \OO''_r} \le |S| f_{\rm LR}(r)
    \end{equation} 
    for a monotonically non-increasing function $f_{\rm LR}$. 
    Furthermore, there exist constants $R>R_1+1$ such that the classical Gibbs state $\pi$ has $(2R_1+2,f)$-SDOC, and
    \begin{subequations}\label{eq:decay_fast0}
        \begin{align}\label{eq:decay_fast}
        f_{\rm LR}(R_1)+(1+f_{\rm LR}(R_1)) f(R-R_1-1) &\le \frac{1}{2} \lr{\frac{1}{c'_d R^d}}^2 \\
        \sum_{r>2R} (r+1)^{d-1}f_{\rm LR}(r-R-1) &\le \frac{1}{8c_d^2 R^d} \label{eq:decay_fast2}
    \end{align}
    \end{subequations}
    for some constant $c'_d>0$ determined by $d$.
    Then the single-site heat-bath Glauber dynamics of $\pi$ has a constant spectral gap \begin{equation}\label{eq:DeltaH>}
        \Delta\ge \lr{\frac{1}{2c_d R^d}}^2 \ee^{-6\cdot 3^dc_d\beta R^d}
    \end{equation} for $\cH$ above the ground state $\kpi$.
\end{thm}

The constant spectral gap implies the Glauber dynamics mixes in $\poly{N}$ time (fast mixing). The form \eqref{eq:LRB} is equivalent to the usual commutator form of Lieb-Robinson bound. For example, for the forward time direction, \begin{align}
    \norm{U^\dagger \OO U - \OO'_r}&\le \int_{\rm Haar} \norm{U^\dagger \OO U- V_{>r}^\dagger U^\dagger \OO U V_{>r}} \mytag{\text{triangle inequality}}\\
    &= \int_{\rm Haar} \norm{[U^\dagger \OO U,V_{>r}]}\nonumber\\
    &\le \max_{V_{>r}}\norm{[U^\dagger \OO U,V_{>r}]}\label{eq:LRB_equiv}
\end{align}
where $\OO'_r=\int_{\rm Haar}V_{>r}^\dagger U^\dagger \OO U V_{>r}$ averages over Haar random unitary $V_{>r}$ supported outside of distance $r$ to $S$.

\eqref{eq:decay_fast0} is satisfied if both $f_{\rm LR}(r),f(r)$ decay sufficiently fast, e.g. faster than $r^{-2d}$. We do not know whether this $2d$ exponent is tight. Nevertheless, we expect it cannot be improved to $r^{-1}$ because of the 3d $\mathbb{Z}_2$ gauge theory example (see the latter Section \ref{sec:gauge} for the model): It is slow mixing \cite{3dgauge_slowmix26} at low temperature, but cannot be SDOC with $f(r)\sim r^{-1}$. This scaling comes from enforcing a flux line entering and exiting the region $C^{\rm c}$ at points $(0,0,\pm r)$, which fluctuates around the origin $A$ in an area $\propto (\sqrt{r})^2$.

Theorem \ref{thm:cohe} requires both unitary connectivity to product state and SDOC. We expect the SDOC condition is necessary: Consider the 3d Ising model at low temperature where the left half of the system has a large uniform longitudinal field favoring the all-zero state. The large field makes the thermal state $\pi$ mainly supported in the ``all-zero'' phase, so it is plausible that the quasi-local $U$ connection holds. However, since the right half does not have the field, we expect a state with all-ones on the right would take exponentially long time to flip, because a large interface between zero and one regions is fairly rigid in the 3d Ising model \cite{statmech_book17}.

Combining with Theorem \ref{thm:cohe}, we have the following Corollary:

\begin{cor}\label{cor:SDOC_gap}
    In the setting of Theorem \ref{thm:cohe}, assume three extra conditions: \begin{enumerate}
        \item The state $\ket{\sqrt{\pi_0}}$ at $s=0$ is a product state.
        \item The $(2,f)$-SDOC decays sufficiently fast \begin{equation}\label{eq:fr<alpha}
        f(r)\le c_\alpha r^{-\alpha}
    \end{equation} with constants $c_\alpha>0,\alpha>3d$ independent of $s$.
        \item The final state $\pi_{1}$ satisfies $(2R_1+2,f_1)$-SDOC with a sufficiently large constant integer $R_1$ determined by $d,c_\alpha,\beta'$, such that there exists \begin{equation}\label{eq:R_R_1}
            R\in \lr{R_1+1, (R_1)^{\frac{\alpha}{3d}} }
        \end{equation} 
        with \begin{equation}\label{eq:f_1<}
            f_1(R-R_1-1) \le \frac{1}{8}\lr{\frac{1}{c'_d R^d}}^2
        \end{equation}
        where $c'_d$ is the constant in \eqref{eq:decay_fast}.
    \end{enumerate}
    Then the single-site Glauber dynamics of $\pi_{1}$ has a constant spectral gap $\Delta>0$ satisfying \eqref{eq:DeltaH>}.
\end{cor}

Comparing to Theorem \ref{thm:cohe}, here we require SDOC with a larger $R_{\rm A}=2R_1+2$. Nevertheless, the correlation \eqref{eq:f_1<} we require is still point-to-set like because $R$ can be asymptotically larger than $R_1$ in \eqref{eq:R_R_1}. Comparing to the SSM condition in Theorem \ref{thm:SSM}, the SDOC here is weaker for the single Gibbs state $\pi_1$. Although it is unclear whether SSM for $\pi_1$ implies the existence of a SDOC path connecting it to a product state, Corollary \ref{cor:SDOC_gap} proves fast mixing for interesting models that are beyond reach by Theorem \ref{thm:SSM}, as we will see.

\begin{proof}
From Theorem \ref{thm:cohe} and product state $\pi_0$, $\ket{\sqrt{\pi_{1}}}=U\ket{\bm{+}}$ with quasi-local unitary $U$ (up to an initial on-site rotation layer) generated by quasi-local Hamiltonian $W=\sum_{u,\ell}W_{u,\ell}$. We arrange the Hamiltonian in a form suitable for applying Lieb-Robinson bounds for power-law interactions: For a given $u$, \begin{equation}\label{eq:suml<}
    \sum_{\ell: 2^\ell \ge \frac{r}{2}} \norm{W_{u,\ell}(s)} \le c_\alpha\beta'\sum_{\ell: 2^\ell \ge \frac{r}{2}} (2^{\ell-1}-1)^{-\alpha} \le c'_\alpha \beta' r^{-\alpha}
\end{equation}
from \eqref{eq:fr<alpha}, for a constant $c'_\alpha$ determined by $\alpha,c_\alpha$. As a result, for any $v,v'\in \Lambda$ with $\mathrm{dist}(v,v')=R$,
\begin{align}
    \sum_{u,\ell:\,W_{u,\ell}\text{ acts on }v,v'} \norm{W_{u,\ell}(s)} &\le \sum_{r=0}^\infty \sum_{u:\mathrm{dist}(u,\{v,v'\})=r} \sum_{\ell:\,W_{u,\ell}\text{ acts on }v,v'} \norm{W_{u,\ell}(s)} \nonumber\\
    &\le \sum_{r=0}^{R/2} \sum_{u:\mathrm{dist}(u,\{v,v'\})=r} \sum_{\ell:\,2^\ell\ge \frac{R}{2}} \norm{W_{u,\ell}(s)} + \sum_{r=R/2}^\infty \sum_{u:\mathrm{dist}(u,\{v,v'\})=r} \sum_{\ell:\,2^\ell\ge r} \norm{W_{u,\ell}(s)} \nonumber\\
    &\le c'_\alpha \beta'\lr{ \sum_{r=0}^{R/2} \sum_{u:\mathrm{dist}(u,\{v,v'\})=r} R^{-\alpha} + \sum_{r=R/2}^\infty \sum_{u:\mathrm{dist}(u,\{v,v'\})=r} r^{-\alpha}} \mytag{\text{Using \eqref{eq:suml<}}} \\
    &\le c_{d,\alpha}\beta' R^{d-\alpha} \label{eq:<d-alpha}
\end{align} 
for a constant $c_{d,\alpha}$ determined by $d,\alpha,c_\alpha$.
Here the first term in the second line uses the fact that if $r\le R/2$, then $u$ is $\ge R/2$ far from one of $v,v'$ so it enforces $2^\ell\ge \frac{R}{2}$. The last line uses \eqref{eq:cd}. 

From \eqref{eq:<d-alpha}, the Lieb-Robinson bound Theorem A.1 in \cite{Hastings_koma06} then implies \eqref{eq:LRB} with \begin{equation}\label{eq:fLR<}
    f_{\rm LR}(r)\le c_{\rm LR}r^{d-\alpha}
\end{equation}
where $c_{\rm LR}$ is a constant determined by $d,\alpha,c_\alpha,\beta'$. Note that the original form of \cite{Hastings_koma06} contains an extra volume factor beyond $|S|$ in \eqref{eq:LRB}, which can be removed by tricks like \eqref{eq:LRB_equiv}; see \cite{LRB_review}.

We then apply Theorem \ref{thm:gap}: Because $f_{\rm LR}$ decays faster than $r^{-2d}$ in \eqref{eq:fLR<}, there exists a sufficiently large constant $R_1$ determined by $d,\alpha,c_\alpha,\beta'$ such that \eqref{eq:decay_fast2} and \begin{equation}\label{eq:fLR<1}
    f_{\rm LR}(R_1) \le \frac{1}{4}\lr{\frac{1}{c'_d R^d}}^2 \le 1
\end{equation}
hold for any $R$ in range \eqref{eq:R_R_1}. Combining \eqref{eq:fLR<1} with assumption \eqref{eq:f_1<}, \eqref{eq:decay_fast} also holds. The spectral gap bound then follows from Theorem \ref{thm:gap}.

\end{proof}

\subsection{Proof of fast mixing}

\begin{proof}[Proof of Theorem \ref{thm:gap}]

Let $B_u$ be the ball centered at $u$ with radius $R-1$. Let $\Pi_u$ be the projector supported in $\overline{B_u}$ projecting onto space \begin{equation}\label{eq:PI_B}
    \mathrm{Span}\lr{\ket{\bz_{\partial B_u}} \otimes \ket{\sqrt{\pi_{B_u}(\cdot|\bz_{\partial B_u})}}}
\end{equation}
i.e. the span of conditional Gibbs states in $B_u$ given any boundary condition on $\partial B_u$.
The Glauber dynamics $\cH$ satisfies the finite-volume Poincare inequality:

\begin{lem}\label{lem:poincare}
For any $u$, \begin{equation}\label{eq:Poincare}
    \sum_{u'\in B_u} \cH_{u'} \ge \Delta_0 (I-\Pi_u),\where \Delta_0 = \min_{v\in \Lambda}\frac{1}{|B_v|}\ee^{-6\cdot 3^d\beta|B_v|}.
\end{equation}
\end{lem}

This is relatively standard, but we provide the proof in Section \ref{sec:poincare} later for self-containedness.

\begin{figure}
    \centering
    \includegraphics[width=0.7\linewidth]{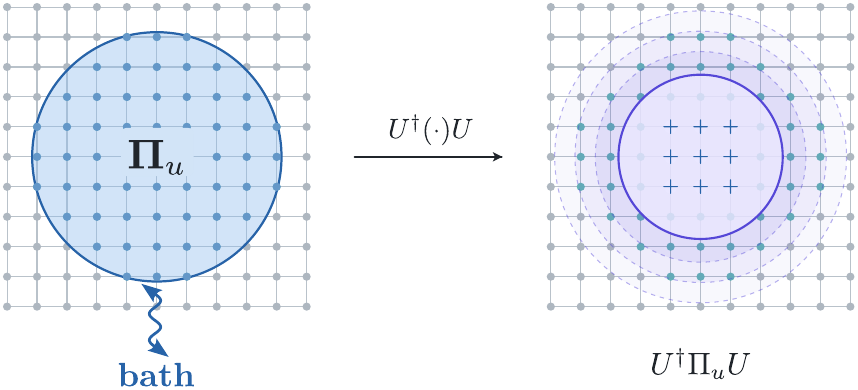}
    \caption{An illustration of the idea of Theorem \ref{thm:gap}: SDOC and $\kpi=U\ket{\bm{+}}$ implies that the interior of the local thermal projector $\Pi_u$ is almost all-$+$ after conjugating by the quasi-local unitary $U$. As a result, any $\ket{-}$ in the conjugated frame triggers constant energy of some $\Pi_u$ from the Poincare inequality \eqref{eq:Poincare}, which leads to a constant spectral gap $\Delta$.}
    \label{fig:Pi}
\end{figure}

We then utilize the fact $\kpi = U\kplus$. The high-level idea of the following proof is illustrated in Fig.~\ref{fig:Pi}.
For each $u$, define projectors $Q_u^\perp=\ket{+}_u\bra{+}$, $Q_u=I-Q_u^\perp$ (those two are single-site) and \begin{equation}\label{eq:Qur=}
    Q_{u,r}:= \lr{\prod_{v: \mathrm{dist}(u,v)\le r-1 } Q_v^\perp}\lr{ I-\prod_{v: \mathrm{dist}(u,v)= r } Q_v^\perp}
\end{equation} 
($r\in \mathbb{Z}_{\ge 0}$) that projects onto states with all-$+$ in the radius-$(r-1)$ ball and at least one not-$+$ in the radius-$r$ ball. $Q_{u,r}=\ket{\bm{+}}\bra{\bm{+}}$ at the smallest $r$ such that the radius-$(r-1)$ ball covers the whole system, and $Q_{u,r}:=0$ for larger $r$. Obviously, $Q_{u,0}=Q_u$ and $I=\sum_{r\ge 0}Q_{u,r}$. We investigate the property of a dressed local ground state projector \begin{equation}\label{eq:tildePI}
    \widetilde{\Pi}_u:=U^\dagger \Pi_u U
\end{equation}
in terms of the subspaces $Q_{u,r}$:

\begin{lem}\label{lem:H>X}
    Given the conditions of Theorem \ref{thm:gap},  \begin{equation}\label{eq:sumr<}
        \sum_{r\ge 0} c_d (r+1)^{d-1}\sup_u\norm{Q_u \widetilde{\Pi}_u Q_{u,r}} \le \frac{1}{2}
    \end{equation}
    where $c_d\ge 1$ is the constant in \eqref{eq:cd}.
\end{lem}

The proof is contained in the latter Section \ref{sec:Pi}, which is the only place in this proof using the SDOC condition.
For any normalized excited state of $\cH$, it is of the form $U\kpsi$ with $\kpsi\perp \kplus$. The energy of $U\kpsi$ is \begin{align}
    \bra{\psi}U^\dagger \cH U\kpsi &= \bra{\psi} U^\dagger \lr{\sum_{u'}\cH_{u'}} U\kpsi \nonumber\\
    &\ge \bra{\psi}U^\dagger \lr{\frac{1}{c_dR^d} \sum_{u\in \Lambda}\sum_{v\in B_u} \cH_{v} }U \kpsi \nonumber\\
    &\ge \frac{\Delta_0}{c_dR^d} \sum_{u\in \Lambda} \bra{\psi}(I-\widetilde{\Pi}_u)\kpsi 
    \label{eq:UHU>DeltaR}
\end{align}
using \eqref{eq:Poincare} and \eqref{eq:tildePI}.
Here the second line holds because \begin{equation}
    \sum_{u'} \cH_{u'} \ge \sum_{u'} a_{u'}\cH_{u'}
\end{equation}
for any set of numbers $a_{u'}\le 1$ due to $\cH_u\ge 0$, and here each $u'$ is contained in at most $c_d R^d$ radius-$R$ balls, so indeed \begin{equation}
    a_{u'}= \frac{1}{c_dR^d} \sum_{u:B_u\ni u'}1\le 1.
\end{equation}

Define
\begin{equation}
    \alpha_r:=\sup_u \norm{Q_u\widetilde{\Pi}_u Q_{u,r}} 
\end{equation}
which is bounded by Lemma \ref{lem:H>X}.
For a fixed $u$, using $I=\sum_r Q_{u,r}$ and the triangle inequality,
\begin{align}
    \norm{Q_u\widetilde{\Pi}_u\kpsi}^2
    &=\norm{\sum_{r\ge 0} Q_u\widetilde{\Pi}_u Q_{u,r}\kpsi}^2 \nonumber\\
    &\le \lr{\sum_{r\ge 0} \alpha_r\norm{Q_{u,r}\kpsi}}^2
\nonumber\\
&\le \lr{\sum_{r\ge 0} \alpha_r}\lr{\sum_{r\ge 0} \alpha_r\norm{Q_{u,r}\kpsi}^2} \mytag{\text{Cauchy-Schwarz inequality}}\\
    &\le \frac12\sum_{r\ge 0} \alpha_r\bra{\psi}Q_{u,r}\kpsi \mytag{\text{Using \eqref{eq:sumr<}}}\\
    &\le \frac12\sum_{r\ge 0} \alpha_r\bra{\psi}\sum_{v:\mathrm{dist}(u,v)=r} Q_v\kpsi 
    \label{eq:QuPiCS}
\end{align}
The last line comes from 
$Q_{u,r}\le \sum_{v:\mathrm{dist}(u,v)=r} Q_v+\ket{\bm{+}}\bra{\bm{+}}$ due to its definition \eqref{eq:Qur=} (i.e. each state in $Q_{u,r}$ has to trigger some $Q_v$ except for the global case $Q_{u,r}=\ket{\bm{+}}\bra{\bm{+}}$).
Since each $v$ is of distance $r$ to at most $c_d(r+1)^{d-1}$ $u$s, summing \eqref{eq:QuPiCS} over $u$ yields
\begin{align}
    \sum_{u\in \Lambda} \norm{Q_u\widetilde{\Pi}_u\kpsi}^2&\le \frac12\sum_v\sum_{r\ge 0}c_d(r+1)^{d-1} \alpha_r\bra{\psi} Q_v\kpsi \nonumber\\
    &\le \frac{1}{4}\bra{\psi} Q\kpsi
    \label{eq:sumQuPi<}
\end{align}
using \eqref{eq:sumr<} and $Q:=\sum_v Q_v$.

On the other hand,
\begin{equation}
    Q_u\kpsi=Q_u(I-\widetilde{\Pi}_u)\kpsi+Q_u\widetilde{\Pi}_u\kpsi .
\end{equation}
Therefore,
\begin{align}
    \bra{\psi}Q\kpsi=\sum_{u\in \Lambda} \norm{Q_u\kpsi}^2 &\le 2\sum_{u\in \Lambda} \norm{Q_u(I-\widetilde{\Pi}_u)\kpsi}^2
        +2\sum_{u\in \Lambda} \norm{Q_u\widetilde{\Pi}_u\kpsi}^2 \nonumber\\
    &\le 2\sum_{u\in \Lambda} \norm{(I-\widetilde{\Pi}_u)\kpsi}^2
        +\frac12\bra{\psi}Q\kpsi 
    \label{eq:Q<G}
\end{align}
using \eqref{eq:sumQuPi<}.
Rearranging and combining with \eqref{eq:UHU>DeltaR} gives
\begin{align}
    \bra{\psi}U^\dagger \cH U\kpsi
    &\ge \frac{\Delta_0}{c_dR^d} \sum_{u\in \Lambda}\norm{(I-\widetilde{\Pi}_u)\kpsi}^2 \nonumber\\
    &\ge \frac{\Delta_0}{c_dR^d}\frac{1}{4} \bra{\psi}Q\kpsi \mytag{\text{Using \eqref{eq:Q<G}}}\\
    &\ge \frac{\Delta_0}{4c_dR^d} \label{eq:H_gap}
\end{align}
where the last line comes from $\bra{\psi} \sum_v Q_v\kpsi\ge 1$ using $\kpsi \perp \kplus$. \eqref{eq:DeltaH>} then follows from \eqref{eq:Poincare} and $|B_u|\le c_d R^d$.

\end{proof}

\subsection{Proof for the dressed ground state projector}\label{sec:Pi}
\begin{proof}[Proof of Lemma \ref{lem:H>X}]

The target \eqref{eq:sumr<} can be separated to two parts:
\begin{align}
    &\quad\sum_{r\ge 0} c_d (r+1)^{d-1} \sup_u\norm{Q_u \widetilde{\Pi}_u Q_{u,r}} \nonumber\\
    &\le \sum_{r= 0}^{2R} c_d (r+1)^{d-1} \sup_u\norm{Q_u \widetilde{\Pi}_u Q_{u,r}} + \sum_{r> 2R} c_d (r+1)^{d-1} \sup_u\norm{Q_u\lr{I-\widetilde{\Pi}_u} Q_{u,r}} \hspace{12em}\mytag{Q_uQ_{u,r>0}=0}\\
    &\le \sum_{r= 0}^{2R} c_d (r+1)^{d-1} \sup_u\norm{Q_u \widetilde{\Pi}_u } + \sum_{r> 2R} c_d (r+1)^{d-1} \sup_u\norm{\lr{I-\widetilde{\Pi}_u} Q_{u,r}} \nonumber\\
    &\le \frac{1}{4}c'_d R^d \sup_u\norm{Q_u \widetilde{\Pi}_u } + \sum_{r> 2R} c_d (r+1)^{d-1} \sup_u\norm{\lr{I-\widetilde{\Pi}_u} Q_{u,r}} \label{eq:sumr<+}
\end{align}
for some constant $c'_d$ determined by $d$.

We first prove \begin{equation}\label{eq:QPI<}
    \norm{Q_u\widetilde{\Pi}_u }\le \frac{1}{c'_d R^d}
\end{equation}
so that the first term in \eqref{eq:sumr<+} is bounded by $1/4$.
Since $Q_u,\widetilde{\Pi}_u$ are projectors, \begin{equation}\label{eq:QPI=}
    \norm{Q_u\widetilde{\Pi}_u }^2=\norm{\widetilde{\Pi}_u Q_u \widetilde{\Pi}_u}=\norm{\Pi_u U Q_u U^\dagger \Pi_u } = \max_{\psi} \alr{\psi|U Q_u U^\dagger|\psi}
\end{equation}
using \eqref{eq:tildePI},
where $\psi$ is maximized over all normalized states with \begin{equation}\label{eq:PIpsi=psi}
    \Pi_u\kpsi=\kpsi.
\end{equation}

Applying Lieb-Robinson bound \eqref{eq:LRB}, $UQ_u U^\dagger$ can be approximated by an operator $\OO'$ supported in the radius-$R_1$ ball $A$ centered at $u$: \begin{equation}\label{eq:LRB'}
    \norm{UQ_u U^\dagger-\OO'}\le \epsilon:=f_{\rm LR}(R_1),\qquad \norm{\OO'}\le \norm{UQ_u U^\dagger}+\epsilon=1+\epsilon.
\end{equation} 
\eqref{eq:QPI=} then becomes \begin{align}
    \norm{Q_u\widetilde{\Pi}_u }^2 &\le \max_{\psi} \alr{\psi|\OO'|\psi} + \epsilon \mytag{\text{Using \eqref{eq:LRB'}}}\\
    &\le \alr{\sqrt{\pi}|\OO'|\sqrt{\pi}} + \max_{\psi}\abs{\alr{\psi|\OO'|\psi}-\alr{\sqrt{\pi}|\OO'|\sqrt{\pi}}}+ \epsilon \nonumber\\
    &\le \mlr{\alr{\sqrt{\pi}|UQ_u U^\dagger|\sqrt{\pi}}+ \epsilon} + \norm{\OO'}\cdot 2f(R-R_1-1) + \epsilon \hspace{20em}\mytag{\text{Using \eqref{eq:LRB'} and SDOC}} \\
    &\le \bra{\bm{+}}Q_u\kplus + 2\epsilon + 2(1+\epsilon)f(R-R_1-1) \mytag{\text{Using \eqref{eq:LRB'} and }\kpi=U\kplus}\\
    &= 2\epsilon + 2(1+\epsilon)f(R-R_1-1) \label{eq:QPI<'}
\end{align}
which implies \eqref{eq:QPI<} from \eqref{eq:decay_fast}. Here the third line uses \begin{equation}
    \alr{\psi|\OO'|\psi}= \alr{\psi|\OO'_{\rm cla}|\psi}
\end{equation}
for any state satisfying \eqref{eq:PIpsi=psi} (including $\kpsi=\kpi$) which can be written as $\kpsi = \sum_{\bz_{\partial A}} \alpha_{\bz_{\partial A}}\ket{\bz_{\partial A}} \otimes \ket{\sqrt{\pi_A(\cdot|\bz_{\partial A})}}\otimes \cdots$ ($\cdots$ is any state outside $\overline{A}=A\sqcup \partial A$), where $\OO'_{\rm cla}$ is a classical (i.e. diagonal) observable \begin{equation}
    \OO'_{\rm cla} = \sum_{\bz_{\partial A}} \bra{\sqrt{\pi_A(\cdot|\bz_{\partial A})}}\OO'\ket{\sqrt{\pi_A(\cdot|\bz_{\partial A})}} \ket{\bz_{\partial A}} \bra{\bz_{\partial A}}
\end{equation}
even if $\OO'$ may contain off-diagonal elements. $\OO'_{\rm cla}$ is supported in $\partial A$ and has $\norm{\OO'_{\rm cla}}\le \norm{\OO'}$. We then applied SDOC to bound the expectation of this classical operator in $\kpsi$ comparing to $\kpi$ in \eqref{eq:QPI<'} following a similar treatment as \eqref{eq:<2f}.

We then bound the second term in \eqref{eq:sumr<+} with $r>2R$. From Lieb-Robinson bound, $\widetilde{\Pi}_u$ can be approximated by an operator $\widetilde{\Pi}_u^{(<r)}$ supported no farther than $r-1$ to $u$, with error 
\begin{equation}\label{eq:Pi_LRB}
    \norm{\widetilde{\Pi}_u-\widetilde{\Pi}_u^{(<r)}}\le |\overline{B_u}| f_{\rm LR}(r-R-1) \le c_d R^d f_{\rm LR}(r-R-1).
\end{equation}
Since $\lr{I-\widetilde{\Pi}_u}\kplus=U^\dagger (I-\Pi_u)U\kplus=0$, \eqref{eq:Pi_LRB} implies \begin{equation}\label{eq:I-PI<r<}
    \norm{\lr{I-\widetilde{\Pi}_u^{(<r)}} \kplus_{<r}}=\norm{\lr{I-\widetilde{\Pi}_u^{(<r)}} \kplus} \le 0+c_d R^d f_{\rm LR}(r-R-1)
\end{equation}
where $\kplus_{<r}$ is the all-$+$ state in the radius-$(r-1)$ ball.
As a result, 
\begin{align}
    \norm{\lr{I-\widetilde{\Pi}_u} Q_{u,r}} &\le \norm{ \lr{I-\widetilde{\Pi}_u^{(<r)}} Q_{u,r} } +  \norm{ \lr{\widetilde{\Pi}_u-\widetilde{\Pi}_u^{(<r)} }Q_{u,r} }\mytag{\text{Triangle inequality}} \\
    &\le \norm{\lr{I-\widetilde{\Pi}_u^{(<r)}}\kplus_{<r}} + \norm{ \lr{\widetilde{\Pi}_u-\widetilde{\Pi}_u^{(<r)} } } \mytag{\text{Using \eqref{eq:Qur=}}} \\
    &\le 2c_d R^d f_{\rm LR}(r-R-1) \label{eq:PiQur<}
\end{align}
using \eqref{eq:Pi_LRB} and \eqref{eq:I-PI<r<}. The second term in \eqref{eq:sumr<+} is then bounded by \begin{equation}\label{eq:sumr>_2ndterm}
    2c_d^2 R^d \sum_{r>2R} (r+1)^{d-1}f_{\rm LR}(r-R-1) \le \frac{1}{4}
\end{equation}
from \eqref{eq:decay_fast2}.

Since both terms in \eqref{eq:sumr<+} are bounded by $1/4$, we get \eqref{eq:sumr<}.

\end{proof}

\subsection{Proof of the finite-volume Poincare inequality}\label{sec:poincare}

\begin{proof}[Proof of Lemma \ref{lem:poincare}]

$\sum_{v\in B_u}\cH_v$ corresponds to the Glauber dynamics on $B_u$ given fixed boundary condition in $\partial B_u$. Since it does not couple spaces with different $\bz_{\partial B_u}$, \eqref{eq:Poincare} is equivalent to for any fixed $\bz_{\partial B_u}$, \begin{equation}\label{eq:H>lambda}
    \bra{\bz_{\partial B_u}}\otimes \bra{\psi}\sum_{v\in B_u}\cH_v\ket{\bz_{\partial B_u}}\otimes \kpsi\ge \Delta_0
\end{equation}
for any normalized $\kpsi\perp \ket{\sqrt{\pi_{B_u}(\cdot|\bz_{\partial B_u})}}$ on $B_u$, i.e. the Glauber dynamics on $B_u$ has spectral gap $\Delta_0$ above its unique steady state $\pi_{B_u}(\cdot|\bz_{\partial B_u})$.
Here the uniqueness comes from the fact that Glauber dynamics on $B_u$ is ergodic in the space of all $\bz_{B_u}$: An ergodic Markov chain always has a unique steady state from the Perron-Frobenius theorem. 

To prove \eqref{eq:H>lambda}, we use the Dirichlet form (hereafter $\bz=\bz_{B_u}$ where we ignore the subscripts) \begin{equation}
    \mathcal{E}(g):=\frac{1}{2}\sum_{\bz,\bz'} \mlr{g(\bz)-g(\bz')}^2 \pi(\bz) P(\bz'\leftarrow \bz)
\end{equation}
for the discrete-time transition matrix $P$ of the Glauber dynamics (where one random site in $B_u$ is resampled at each timestep), and
apply Lemma 13.18 in textbook \cite{levin_markovchains}:

\begin{lem}[Adapted from \cite{levin_markovchains}]
    For two detailed-balanced transition matrices $P,\widetilde{P}$ with steady states $\pi,\widetilde{\pi}$ respectively, if $\widetilde{\mathcal{E}}(g)\le \alpha \mathcal{E}(g)$ for all function $g$, then their spectral gaps satisfy \begin{equation}\label{eq:Delta>Delta}
        \Delta \ge \frac{1}{\alpha} \widetilde{\Delta}\cdot \min_{\bz} \frac{\widetilde{\pi}(\bz)}{\pi(\bz)} .
    \end{equation}
\end{lem}

We choose $\widetilde{P}$ to be the infinite-temperature Glauber dynamics (i.e. independent bit-flips) with $\widetilde{\pi}\propto I$ and $\widetilde{\Delta}=1/|B_u|$. Since \begin{equation}
    \frac{\pi(\bz)}{\widetilde{\pi}(\bz)}\le \ee^{2\cdot 3^d\beta |B_u|},\qquad \frac{\widetilde{\pi}(\bz)\widetilde{P}(\bz'\leftarrow \bz)}{\pi(\bz)P(\bz'\leftarrow \bz)} \le \ee^{4\cdot 3^d\beta |B_u|}
\end{equation}
from the similar energy bound argument as \eqref{eq:piA>}, 
\begin{equation}
    \widetilde{\mathcal{E}}(g)\le \ee^{4\cdot 3^d\beta |B_u|} \mathcal{E}(g)
\end{equation}
so \eqref{eq:Delta>Delta} leads to \eqref{eq:H>lambda} with \begin{equation}
    \Delta_0 = \ee^{-2\cdot 3^d\beta |B_u|}\cdot \frac{1}{|B_u|}\cdot \ee^{-4\cdot 3^d\beta |B_u|} = \frac{1}{|B_u|}\ee^{-6\cdot 3^d\beta |B_u|}.
\end{equation}
One can then minimize over $u$ so that $\Delta_0$ becomes a constant independent of $u$.

\end{proof}

\section{Applications}\label{sec:applications}

We then apply our results to various classical models. 

\subsection{Quasi-local connection at sufficiently high temperature or large field}

First, Theorem \ref{thm:cohe} directly applies to any model satisfying SSM, which include two general limits: \begin{enumerate}
    \item SSM is satisfied for any model $H$ at sufficiently high temperature $\beta<\beta_{\rm high}$ with a constant $\beta_{\rm high}>0$ determined by $d$.
    \item For any model $H$, consider adding a strong on-site field \begin{equation}\label{eq:H+field}
        H'= H - \beta h\sum_u \mathbb{I}[z_u=0]
    \end{equation}
    that favors the $0$ state. Note that \eqref{eq:H+field} is just an example because the field can be more general that differentiates the other states. SSM is satisfied if $h>h_{\rm large}$ for a constant $h_{\rm large}>0$ determined by $d,q$.
\end{enumerate}
These two regimes obey SSM because they satisfy the Dobrushin's condition
of weak dependence \cite{Dobruschin_state68,statmech_book17}. For the strong-field case for example, the intuition is that each site strongly favors state $0$ no matter what its neighbors are, so any dependence among the sites cannot percolate.

Since SSM directly implies rapid mixing for Glauber dynamics \cite{SSM_rev04}, our spectral gap Theorem \ref{thm:gap} would not yield new results. A quasi-local unitary connection between coherent Gibbs states \eqref{eq:kpi=Ukpi} is also known, because they are unique ground states of a gapped path of Hamiltonians $\cH$ in \eqref{eq:H=L} so \eqref{eq:kpi=Ukpi} follows from quasi-adiabatic continuation \cite{quasiadiab05,quasiadiab12}. Nevertheless, the quasi-local tail of $W$ from quasi-adiabatic continuation would be a stretched exponential, which can be improved to exponential by our results. Moreover, we obtain the exact quasi-local Lindbladian connection \eqref{eq:2way} from Theorem \ref{thm:cohe}, which improves upon approximate $\mathrm{polylog}N$-time or $\mathrm{polylog}N$-locality dynamics from rapid-mixing or channel circuits \cite{thermal_prepare_cor_BrandaoKastoryano19,thermal_approxMarkov19,thermal_circuit26}.

\begin{cor}\label{cor:highT}
    For any classical $H$ with $\beta<\beta_{\rm high}$, its Gibbs state satisfies
    \begin{equation}
        \kpi = \mathcal{T} \ee^{-\ii \int^{1}_{0} W(s) \dd s} \ket{\bm{+}}
    \end{equation}
    and \begin{equation}\label{eq:pi=I}
        \pi = \mathcal{T}\ee^{\int^{1}_{0} \cL^{1\leftarrow 0}(s) \dd s} [\hat{I}]
    \end{equation}
    with maximally mixed state $\hat{I}$, where $W(s)=\sum_{u,r}W_{u,r}(s),\cL^{1\leftarrow 0}(s)=\sum_{u,r}\cL^{1\leftarrow 0}_{u,r}(s)$ are exponentially quasi-local in the sense that $W_{u,r}(s),\cL^{1\leftarrow 0}_{u,r}$ are supported in the ball of radius $r$ centered at $u\in \Lambda$, and \begin{equation}\label{eq:Lur<}
        \norm{W_{u,r}(s)},\norm{\cL^{1\leftarrow 0}_{u,r}(s)}_\diamond \le c_1 \ee^{-c_2 r}
    \end{equation}
    for constants $c_1,c_2>0$ determined by $d$. The same conclusion holds for the Gibbs state of \eqref{eq:H+field} at large field $h>h_{\rm large}$, except for an extra $\beta h$ dependence in $c_1,c_2>0$.
\end{cor}

Here we only show one direction in \eqref{eq:pi=I} because any state $\pi$ can evolve to $\hat{I}$ by on-site depolarization channels.

\begin{proof}
    We first prove the high temperature case. $\pi$ satisfies SSM, and therefore $(2,f)$-SDOC with an exponential function $f=c_1'\ee^{-c_2'r}$ for constants $c_1',c_2'$ determined by $d$. Given $H$, choose a smooth path $\pi_s$ of Gibbs states corresponding to $H_s=sH$ in $s\in[0,1]$, which is always in high temperature regime and satisfies $(2,f)$-SDOC uniformly. The statement then follows from Theorem \ref{thm:cohe}.

    The strong field case follows similarly, where the smooth path is instead $H_s = H'-(1-s)H$ that gradually turns on/off the $H$ part. The Gibbs state of $H_1=H'$ is then two-way connected to the product state at $s=0$, which is further connected to the maximally mixed state by on-site channels. The coherent Gibbs state is also connected to $\ket{\bm{+}}$ by extra on-site rotations.
\end{proof}

\subsection{Quasi-local connection of the Ising model with an external field}

From now on we focus on specific models, where results can be shown beyond the perturbative regimes in Corollary \ref{cor:highT}.

The Ising model \eqref{eq:Ising} has a well understood phase diagram in Fig.~\ref{fig:Potts}(a): There is a critical $\beta_{\rm c}^{\rm Ising}$ determined by $d$ ($\beta_{\rm c}^{\rm Ising}=+\infty$ at $d=1$) such that the line $(\beta>\beta_{\rm c}^{\rm Ising}, h=0)$ is a spontaneous-symmetry-breaking (SSB) phase, while the rest of the plane excluding the critical point $(\beta_{\rm c}^{\rm Ising}, h=0)$ is in the trivial phase \cite{Ising_DLR72,statmech_book17}. Mathematically, the two phases can be diagnosed by the Dobrushin-Lanford-Ruelle (DLR) Gibbs state in the thermodynamic limit \cite{Dobruschin_state68,Dobruschin_state69}: The trivial phase has a unique Gibbs state when boundaries are taken to be infinitely far, while the symmetry-breaking phase has more than one Gibbs states. Another diagnostic is the analyticity of the free energy density $N^{-1}\log \mathcal{Z}$ \cite{leeyangzero}. It is an analytic function of $\beta$ in the the symmetry-breaking phase, and of both $\beta,h$ in the trivial phase \cite{Ising_WSM_nonzeroh19,Ising_analytic20}.

Here we provide another diagnostic of the two phases that improves the $\mathrm{polylog}(N)$-depth circuits in \cite{thermal_circuit26}.
Since the trivial phase satisfies exponential WSM \cite{Ising_WSM_nonzeroh19}, Theorem \ref{thm:cohe} implies quasi-local connection in the whole phase following the proof of Corollary \ref{cor:highT}.

\begin{cor}\label{cor:Ising}
    For any $(\beta,h)$ where either $h\neq 0$ or $\beta<\beta_{\rm c}^{\rm Ising}$, the Ising model Gibbs state $\pi$ satisfies the same conclusions in Corollary \ref{cor:highT} for constants $c_1,c_2>0$ determined by $d,\beta,h$.
\end{cor}

As mentioned earlier, the 3d Ising model is believed to violate SSM at large $\beta$ and small $h>0$ \cite{Martinelli99}.
Nevertheless, the Glauber dynamics $\cL$ is known to be rapid mixing by combining WSM and monotonicity of the model \cite{Martinelli99}, which covers the implication from Theorem \ref{thm:gap}. More precisely, the monotonicity property refers to \cite{Martinelli99}\begin{equation}\label{eq:monotonic}
    (-1)^{z_u}\bra{\bm{z}(u)}\cL\ket{\bm{z}} \le (-1)^{z_u}\bra{\bm{z}'(u)}\cL\ket{\bm{z}'},\qquad \forall \bm{z}\le \bm{z}',u\in \Lambda \text{ such that } z_u=z'_u
\end{equation}
where $\bm{z}(u)$ is bitstring $\bm{z}$ with $z_u$ flipped, and $\bm{z}\le \bm{z}'$ means the partial order $z_u\le z'_u (\forall u\in \Lambda)$, i.e. $\bm{z}'$ can be obtained by flipping some $z_u=0$ to $1$ in $\bm{z}$. For the case $z_u=z'_u=0$, \eqref{eq:monotonic} means that $u$ in state $\bm{z}'$ is more likely to be flipped to $1$ comparing to state $\bm{z}$, because $\bm{z}'$ has ``more'' $1$s. The case $z_u=z'_u=1$ is similar, where $u$ in state $\bm{z}'$ is less likely to be flipped to $0$. The Glauber dynamics of the Ising model is monotonic, because the contribution from external field $h$ cancels from both sides of \eqref{eq:monotonic}, while the ferromagnetic Ising interaction tends to align the variables.

\subsection{The Potts model above the critical temperature}

Similar to the Ising model, the Potts model \eqref{eq:Potts} has a critical temperature $\beta^{\rm Potts}_{\rm c}$ determined by $d,q$ ($\beta_{\rm c}^{\rm Potts}=+\infty$ at $d=1$) separating the high-temperature trivial phase $\beta<\beta_{\rm c}$ and the low-temperature symmetry-breaking phase $\beta>\beta_{\rm c}$, diagnosed by whether the DLR Gibbs state is unique \cite{Potts_DLR87,Potts_DLR01,Potts_book06}. The free energy density is also analytic in $\beta$ in the trivial phase \cite{Ising_analytic20}. It is not known however whether it is analytic in the whole low temperature phase; there analyticity is established in special cases of (\emph{1}) the Ising case $q=2$ above, (\emph{2}) sufficiently low temperature $\beta>\beta_{\rm large}$ for a constant $\beta_{\rm large}>\beta^{\rm Potts}_{\rm c}$ \cite{potts_superlowT_analy06}, and (\emph{3}) $d=2$ where the two phases are dual to each other from a generalization of Kramers-Wannier duality \cite{Potts_2d_Tc13}, so analyticity in the high-temperature phase suffices \cite{Ising_analytic20}.

The Potts model satisfies exponential WSM in the whole high temperature phase \cite{Potts_WSM20} building on results of correlation decay \cite{Potts_sharp19}. The Potts model satisfies SSM in the whole high temperature phase for $d=2$ \cite{WSMSSM_2d94} and for $q=2$, but it is believed that SSM is violated in an intermediate temperature range $(\beta_{\rm SSM},\beta^{\rm Potts}_{\rm c})$ with constant $\beta_{\rm SSM}<\beta^{\rm Potts}_{\rm c}$ for $d,q\ge 3$ \cite{Potts_randcluster_mixing24}. Nevertheless, our results imply quasi-local connection and fast mixing throughout the high temperature phase, illustrated in Fig.~\ref{fig:Potts}(b).

\begin{cor}\label{cor:potts}
    For any $q\ge 2$ and $\beta<\beta^{\rm Potts}_{\rm c}$, the Potts model Gibbs state $\pi$ satisfies the same conclusions in Corollary \ref{cor:highT} with different constants $c_1,c_2>0$ determined by $d,q,\beta$. Furthermore, the Glauber dynamics has a spectral gap at least $\Delta>0$ determined by $d,q,\beta$, and is therefore fast mixing.
\end{cor}

\begin{proof}
    The quasi-local connection part is the same as the proof of Corollary \ref{cor:highT}, where the whole trivial phase satisfies exponential WSM from the proof of Corollary 1.5 in \cite{Potts_WSM20}. The conclusion on Glauber dynamics follows similarly from Corollary \ref{cor:SDOC_gap}, where an exponential-decaying $f$ satisfies \eqref{eq:fr<alpha} for $\alpha=3d+1$ for example.
\end{proof}

\subsection{The Ising model with frustrated perturbations at sufficiently low temperature}

The Ising and Potts phases discussed above where our results apply include the infinite temperature maximally mixed state. In this subsection and the next we discuss phases that naturally emerge at sufficiently low temperature, where our results could yield meaningful predictions.

\begin{figure}
    \centering
    \includegraphics[width=0.8\linewidth]{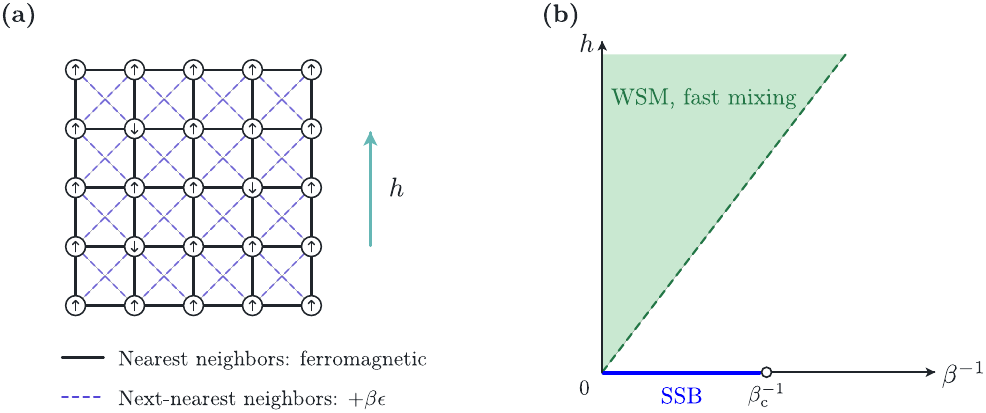}
    \caption{(a) The Ising model perturbed by next-nearest-neighbor anti-ferromagnetic couplings, which inhibits fast mixing proofs based on monotonicity. (b) Its phase diagram at sufficiently small $\epsilon<\epsilon_{\rm c}$, where we are able to prove WSM and fast mixing in the green region, together with quasi-local connections. Whether these results generalize to the whole plane except the SSB segment remains a conjecture.}
    \label{fig:frust}
\end{figure}

One example is the Ising model \eqref{eq:Ising} at nonzero $h$ weakly perturbed by frustrated interactions:
\begin{equation}\label{eq:Ising_frust}
 H^{\rm frust} = H^{\rm Ising} + \beta\epsilon \sum_{\alr{u,v}_{\rm nnn}} (1-2z_u)(1-2z_v)
\end{equation}
where $\alr{u,v}_{\rm nnn}$ sums over unordered next-nearest-neighbors pairs $u,v$ in the square lattice, i.e. pairs with Euclidean distance $\sqrt{2}\times$ lattice spacing; see Fig.~\ref{fig:frust}(a). \eqref{eq:Ising_frust} is just an example; one can verify that the following discussions generalize to any $\mathbb{Z}_2$-symmetric local perturbations. We assume $h>0$ that favors the all-zero state $\bm{0}$ without loss of generality. If $\epsilon>0$ is sufficiently small comparing to $1$ but not necessarily to $h$, one expects the model is in the same trivial phase as the unperturbed one. However, the perturbation \emph{destroys monotonicity} \eqref{eq:monotonic} of the Glauber dynamics: Consider $z_u=z'_u=0$ for example, if $\bm{z},\bm{z}'$ only differs at a next-nearest-neighbor $v$ of $u$ with $z_v=0,z'_v=1$, then $u$ is more likely to be flipped in the case of $\bm{z}$ instead. The frustrated model is also expected to violate SSM for certain parameters $(h,\epsilon)$ as the unperturbed model, because the boundary order mechanism in Section \ref{sec:SSM} still applies. Therefore, previous techniques are insufficient to deduce fast mixing properties of the Glauber dynamics beyond the large $h$ limit in Corollary \ref{cor:highT}.

Here, we are able to show fast mixing along with quasi-local connections as long as the field is large comparing to temperature, which includes cases with $h\ll 1$. See Fig.~\ref{fig:frust}(b) for the phase diagram.

\begin{cor}\label{cor:Ising_frustrated}
    If $0\le \epsilon<\epsilon_{\rm c}$ and $\beta h>c_{\rm frust}$ for constants $\epsilon_{\rm c},c_{\rm frust}$ determined by $d$, the Gibbs state of \eqref{eq:Ising_frust} satisfies the same conclusions in Corollary \ref{cor:potts} with different constants $c_1,c_2,\Delta>0$ determined by $d,h,\beta$.
\end{cor}

\begin{proof}
The equilibrium properties at large $\beta h$ can be controlled by the Pirogov-Sinai theory \cite{pirogovsinai75,pirogovsinai_Zahradnik84,statmech_book17}. In particular, we can directly apply \cite{PStheory_trivialphase97} (see also \cite{pirogovsinai_unique06}) for our case that has a unique translation invariant ground state. Nevertheless, here we present a more elementary route that does not depend on explicit translation invariance properties or cluster expansion techniques involved in Pirogov-Sinai theory. 

The only condition we need for the model is a \emph{Peierls condition} for the Hamiltonian: For any bitstring $\bz$, \begin{equation}\label{eq:peierls}
        H^{\rm frust}(\bz)-H^{\rm frust}(\bm{0}) \ge \beta h|\bz|
    \end{equation}
    where $|\bz|$ is the number of ones in $\bz$. To see \eqref{eq:peierls}, its left hand side contains three contributions from \eqref{eq:Ising_frust} and \eqref{eq:Ising}: \begin{enumerate}
        \item The field term contributes $\beta h|\bz|$.
        \item The ferromagnetic coupling contributes $\beta \sum_{\alr{u,v}}\mathbb{I}[z_u\neq z_v]$.
        \item The antiferromagnetic perturbation contributes \begin{align}
            -2\beta\epsilon \sum_{\alr{u,v}_{\rm nnn}}\mathbb{I}[z_u\neq z_v] &\ge -2\beta\epsilon \sum_{\alr{u,v}_{\rm nnn}} \lr{\mathbb{I}[z_u\neq z_{w}]+\mathbb{I}[z_{w}\neq z_v]} \nonumber\\
            &\ge -\frac{\epsilon}{\epsilon_{\rm c}}\beta \sum_{\alr{u,v}}\mathbb{I}[z_u\neq z_v]\label{eq:antiferro<ferro}
        \end{align}
        where is $w$ the middle vertex between $u,v$ such that $w,u$ and $w,v$ are neighbor pairs; we just choose one $w$ if there are multiple choices. The first line of \eqref{eq:antiferro<ferro} uses the fact that if $z_u\neq z_v$, then at least one of the pairs $w,u$ or $w,v$ should not agree. The last line of \eqref{eq:antiferro<ferro} holds with a constant $\epsilon_{\rm c}$ determined by $d$, because each nearest-neighbor pair appears in only constant number of terms in the first line.
    \end{enumerate}
Combining the three contributions then lead to \eqref{eq:peierls} for $\epsilon<\epsilon_{\rm c}$, because the antiferromagnetic one is dominated by the ferromagnetic one, which is nonnegative.

We then utilize the Peierls condition to prove SDOC. Actually we are able to show WSM. 
Consider regions $A,C$ with distance $r$ and a fixed bitstring $\bz_C$ in the setting of Definition \ref{def:SDOC}, we are interested in the conditional probability $\pi\lr{\bz_{C^{\rm c}}|\bz_C}$ where $\pi$ is the Gibbs state of $H^{\rm frust}$. Each $\bz_{C^{\rm c}}$ is equivalently described by the set $S\subset C^{\rm c}$ containing state $1$. We consider two copies of the state $\pi_{C^{\rm c}}\lr{\cdot|\bz_C}\otimes \pi_{C^{\rm c}}\lr{\cdot|\bz'_C}$ with different boundary conditions, and randomly draw $(S,S')$ from it corresponding to $\bz_{C^{\rm c}},\bz'_{C^{\rm c}}$. We denote the event $A\leftrightarrow C$ to be the case where $S\cup S'$ connects $A$ to $C$, i.e. there is a connected subset $R\subset S\cup S'$ that overlaps with $A$ and connects to $C$: $\mathrm{dist}(R,C)=1$. See Fig.~\ref{fig:2copy}(a) for an illustration. We also let $A\not\leftrightarrow C$ be the complementary event. We claim that \begin{equation}\label{eq:pi-pi<AC}
    \norm{\pi_A(\cdot|\bz_C)-\pi_A(\cdot|\bz'_C)}_1\le 2\mathbb{P}[A\leftrightarrow C]
\end{equation}
bounded by the probability of the connecting event. The reason is as follows. For any observable $\OO_A$ in $A$ in the first copy with $\norm{\OO_A}=1$, let $\OO'_A$ be the same observable but in the second copy. For the $A\not\leftrightarrow C$ case, $S\cup S'=R_A\sqcup R_C$ can be separated to $R_A,R_C$ that do not connect with each other, such that $R_C$ does not overlap with $A$ and $R_A$ does not connect with $C$. This is shown in Fig.~\ref{fig:2copy}(b). We then have expectation 
\begin{align}
    \mathbb{E}[(\OO_A-\OO'_A)\mathbb{I}[A\not\leftrightarrow C]] &= \sum_{\bz_{C^{\rm c}},\bz'_{C^{\rm c}}:A\not\leftrightarrow C} (\OO_A(\bz)-\OO'_A(\bz'))\cdot\pi(\bz|\bz_C)\pi(\bz'|\bz'_C)\nonumber\\
    &= \sum_{\bz_{C^{\rm c}},\bz'_{C^{\rm c}}:A\not\leftrightarrow C} (\OO_A(\widetilde{\bz}')-\OO'_A(\widetilde{\bz}))\cdot\pi(\bz|\bz_C)\pi(\bz'|\bz'_C) \nonumber\\
    &= \sum_{\widetilde{\bz}_{C^{\rm c}},\widetilde{\bz}'_{C^{\rm c}}:A\not\leftrightarrow C} (\OO_A(\widetilde{\bz}')-\OO'_A(\widetilde{\bz})) \cdot\pi(\widetilde{\bz}|\bz_C)\pi(\widetilde{\bz}'|\bz'_C)\nonumber\\
    &=-\mathbb{E}[(\OO_A-\OO'_A)\mathbb{I}[A\not\leftrightarrow C]] \nonumber\\
    &= 0. \label{eq:AnotC}
\end{align}
Here in the second line, we have defined $\widetilde{\bz},\widetilde{\bz}'$
to be $\bz,\bz'$ but with the bits in $R_A$ exchanged between the two copies, i.e. $\widetilde{z}_u=z'_u$ ($z_u$) for all $u\in R_A$ ($u\not\in R_A$). This leads to e.g. $\OO_A(\widetilde{\bz})=\OO_A(\bz')$ which we have used. The third line of \eqref{eq:AnotC} follows because (\emph{1}) the exchange map is one-to-one in the space of bitstrings satisfying $A\not\leftrightarrow C$, and (\emph{2}) the product probability factorizes into a product of contributions from $R_A$ and $R_C$ due to locality of $H$, while exchanging bits in $R_A$ does not change the probability because the $R_A$ contributions are independent of the boundary condition. Since \eqref{eq:AnotC} vanishes, 
\begin{equation}
    \abs{\pi(\OO_A|\bz_C)-\pi(\OO_A|\bz'_C)}=\abs{\mathbb{E}[\OO_A-\OO'_A]}\le \mathbb{E}[\abs{\OO_A-\OO'_A}\mathbb{I}[A\leftrightarrow C]] \le 2 \mathbb{P}[A\leftrightarrow C]
\end{equation}
using $\norm{\OO_A}=1$,
which implies \eqref{eq:pi-pi<AC}.

\begin{figure}
    \centering
    \includegraphics[width=0.95\linewidth]{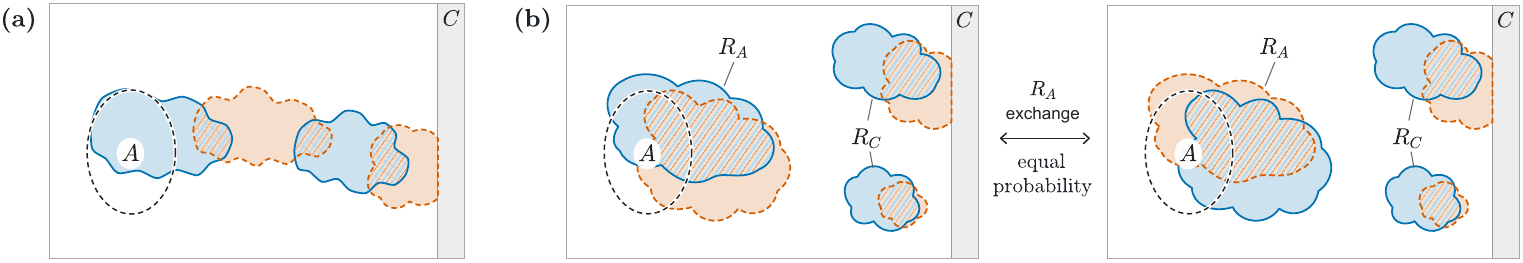}
    \caption{An illustration for the proof of Corollary \ref{cor:Ising_frustrated}, where the Peierls condition implies SDOC. (a) A configuration of the two copies such that $A\leftrightarrow C$. Here blue (orange) regions denote $1$s in the first (second) copy. (b) A configuration with $A\not\leftrightarrow C$. Swapping the $R_A$ regions between the two copies preserve the joint probability, but flips the sign of observable $\OO_A-\OO'_A$. }
    \label{fig:2copy}
\end{figure}

We then bound the probability of the connecting event from a Peierls argument. First, \begin{equation}\label{eq:P<PuC}
    \mathbb{P}[A\leftrightarrow C] \le \sum_{u\in A} \mathbb{P}[u\leftrightarrow C]
\end{equation}
from the union bound,
where $u\leftrightarrow C$ is defined similarly as the event that $S\cup S'\ni u$ connects to $C$.
For any $S,S'$ such that $u\leftrightarrow C$, there exists a unique connected component $R$ of $S\cup S'$ that connects $u$ to $C$. Moreover, as long as $r>c_{\rm r}$ for a constant $c_{\rm r}$ determined by $d, h$, there exists a distance 
$\rho\in( r/2,r-2]$ such that
\begin{equation}\label{eq:R-R<R}
    \abs{R_{\rho+1} }-\abs{R_\rho} \le c'_{\rm r}h\abs{R_\rho} 
\end{equation}
where $R_\rho\subset R$ is the vertices of $R$ within $R$-distance $\le \rho$ to $u$, and $R$-distance means the length of the shortest path contained in $R$ that connects the vertex to $u$; $R_\rho$ defined in this way is still connected. The constant $c'_{\rm r}$ can be made arbitrarily small by enlarging the constant $c_{\rm r}$. Indeed, if \eqref{eq:R-R<R} does not hold for any $\rho\ge r/2$, then it would yield an exponentially large volume \begin{equation}
    |R_r| \ge \lr{1+c'_{\rm r}h}^{\frac{r}{2}-2}
\end{equation}which contradicts with $|R_r|\le c_d r^d$ from \eqref{eq:cd}. Therefore \eqref{eq:R-R<R} has to hold for some 
$\rho\in( r/2,r-2]$.

We view $R_\rho$ as a fundamental object, where each given $R_\rho$ may correspond to different choices of $R$. More generally, we say $(\bz_{C^{\rm c}},\bz'_{C^{\rm c}})$ \emph{contains} $R_\rho$ as long as $S\cup S'\supset R_\rho$ and $\abs{(S\cup S')\cap \partial R_\rho}\le c'_{\rm r}h|R_\rho|$; they do not even need to connect $u$ to $C$. For any $(\bz_{C^{\rm c}},\bz'_{C^{\rm c}})$ containing $R_\rho$, we compare it with a modified state $(\bz^{\rm e}_{C^{\rm c}},{\bz'}^{\rm e}_{C^{\rm c}})$ corresponding to $(S\setminus R_\rho, S'\setminus R_\rho)$, i.e. \emph{erasing} $1$s in $R_\rho$ to $0$s. We have 
\begin{equation}\label{eq:z_vs_flip}
    \frac{\pi(\bz_{C^{\rm c}}|\bz_C)}{\pi(\bz^{\rm e}_{C^{\rm c}}|\bz_C)} \le \ee^{-\beta h|R_\rho \cap S|}\cdot \ee^{c''_dc'_{\rm r}\beta h|R_\rho|}
\end{equation}
for a constant $c''_d$ determined by $d$; a similar bound holds for the second copy. \eqref{eq:z_vs_flip} just follows from \eqref{eq:peierls} on erasing a $1$-region with all-zero boundaries; here the boundary is not all-zero, whose energy difference is bounded by the boundary size giving rise to the second factor in \eqref{eq:z_vs_flip}. Multiplying \eqref{eq:z_vs_flip} with the second copy yields 
\begin{equation}\label{eq:z_vs_flip1}
    \frac{\pi(\bz_{C^{\rm c}}|\bz_C)}{\pi(\bz^{\rm e}_{C^{\rm c}}|\bz_C)}\cdot \frac{\pi(\bz'_{C^{\rm c}}|\bz'_C)}{\pi({\bz'}^{\rm e}_{C^{\rm c}}|\bz'_C)} \le \ee^{-\beta h|R_\rho|}\cdot \ee^{2c''_dc'_{\rm r}\beta h\abs{R_\rho}} \le \ee^{-\frac{1}{2}\beta h|R_\rho|}
\end{equation}
using $|R_\rho \cap S|+|R_\rho \cap S'|\ge |R_\rho|$ and \eqref{eq:R-R<R} with a sufficiently small constant $c'_{\rm r}$ determined by $d$.
As in the Peierls argument, we sum over bitstrings $\bz_{C^{\rm c}},\bz'_{C^{\rm c}}$ containing $R_\rho$:
\begin{align}
    \sum_{\bz_{C^{\rm c}},\bz'_{C^{\rm c}}:\text{ contains } R_\rho} \pi(\bz_{C^{\rm c}}|\bz_C)\pi(\bz'_{C^{\rm c}}|\bz'_C) &\le \ee^{-\frac{1}{2}\beta h|R_\rho|}\sum_{\bz_{C^{\rm c}},\bz'_{C^{\rm c}}: \text{ contains } R_\rho} \pi(\bz^{\rm e}_{C^{\rm c}}|\bz_C)\pi({\bz'}^{\rm e}_{C^{\rm c}}|\bz'_C) \nonumber\\
    &\le \ee^{c_{\rm P}|R_\rho|} \ee^{-\frac{1}{2}\beta h|R_\rho|} \label{eq:probR<}
\end{align}
for a constant $c_{\rm P}$ determined by $d$.
Here we have used the fact that each $(\bz^{\rm e}_{C^{\rm c}},{\bz'}^{\rm e}_{C^{\rm c}})$ corresponds to at most $c_{\rm P}^{|R_\rho|}$ different $(\bz_{C^{\rm c}},\bz'_{C^{\rm c}})$, because two pairs of $(\bz_{C^{\rm c}},\bz'_{C^{\rm c}})$ that differ outside $R_\rho$ map to different $(\bz^{\rm e}_{C^{\rm c}},{\bz'}^{\rm e}_{C^{\rm c}})$; the choices inside $R_\rho$ are however only exponentially many in its volume. 
Since each valid $R_\rho$ has large volume 
$|R_\rho|\ge r/2$, \eqref{eq:probR<} and the union bound imply \begin{equation}\label{eq:PuC<}
    \mathbb{P}[u\leftrightarrow C] \le \sum_{\text{connected }R_\rho\ni u:|R_\rho|\ge r/2} \ee^{c_{\rm P}|R_\rho|} \ee^{-\frac{1}{2}\beta h|R_\rho|} \le \ee^{-c_{\rm P}'r}
\end{equation}
as long as $\beta h>c_{\rm frust}$ with constants $c'_{\rm P},c_{\rm frust}$ determined by $d$. \eqref{eq:PuC<} converges because there are at most exponentially many connected subsets at a given size containing a given vertex $u$, in any bounded degree graph. Combining \eqref{eq:pi-pi<AC}, \eqref{eq:P<PuC} and \eqref{eq:PuC<}, 
\begin{equation}
    \norm{\pi_A(\cdot|\bz_C)-\pi_A(\cdot|\bz'_C)}_1 \le 2|A|\ee^{c_{\rm P}'c_{\rm r}}\ee^{-c_{\rm P}'r},
\end{equation}so the model is 
$2\ee^{c_{\rm P}'c_{\rm r}}\ee^{-c_{\rm P}'r}$-WSM. Note that this conclusion is stronger than the one in \cite{PStheory_trivialphase97} that requires $r$ to be at least at the order of $\mathrm{diam}(A)$.

One can then follow the proof of Corollaries \ref{cor:highT} and \ref{cor:potts} to conclude Corollary \ref{cor:Ising_frustrated}. In particular, we consider the smooth path $H_s = sH^{\rm frust}+ (1-s)\beta h \sum_u z_u $ that connects to the trivial state with field-only $H_0$. Along this path, the Peierls condition \ref{eq:peierls} holds uniformly, so as the WSM above.

\end{proof}

\subsection{The 3d $\mathbb{Z}_2$ gauge theory with an external field at sufficiently low temperature}\label{sec:gauge}

\begin{figure}
    \centering
    \includegraphics[width=0.95\linewidth]{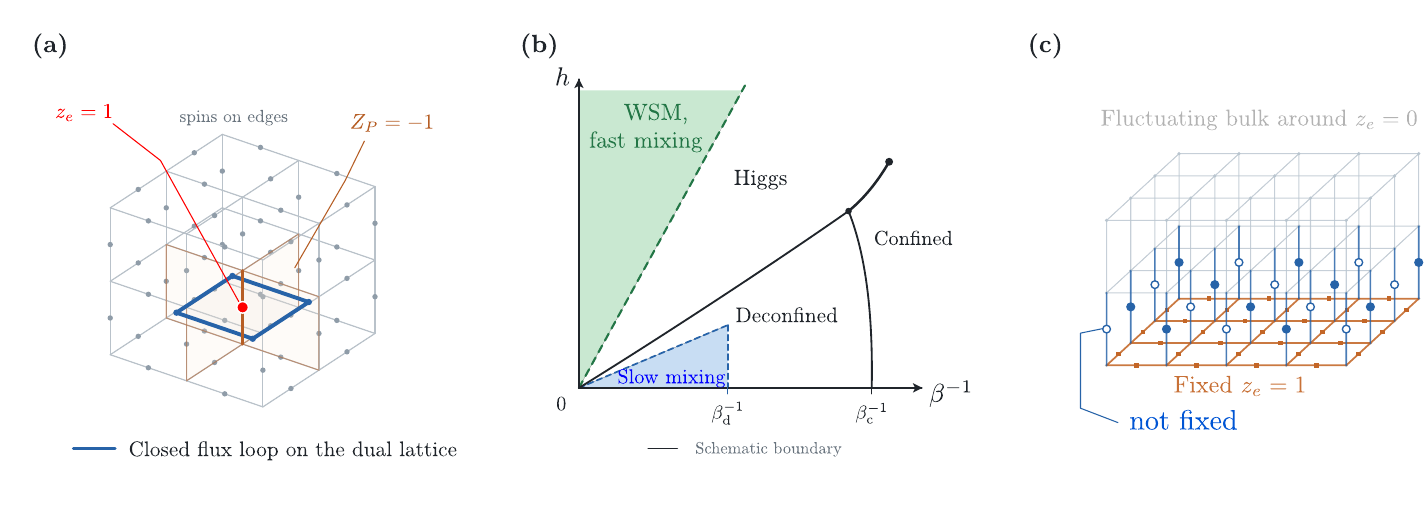}
    \caption{The 3d $\mathbb{Z}_2$ gauge theory with a field. (a) The gauge theory Hamiltonian is composed of plaquette $Z_P$s, whose excitations form closed loops. Here a single $z_e=1$ in the $0$ background induces a smallest loop. (b) The phase diagram where the blue region is proven to be slow mixing \cite{3dgauge_slowmix26}, and our results apply to the green region. (c) A candidate boundary condition (yellow) obstructing SSM. At low temperature, the bulk (gray) tends to be at $z_e=0$, so the fixed $z_e=1$ boundaries induces a 2d layer of anti-ferromagnetic Ising interactions among the blue edges, which potentially gives rise to a $\mathbb{Z}_2$ boundary order that violates SSM. More precisely, the figures shows one of the two checkerboard patterns on the blue edges, which cannot be distinguished by the field. }
    \label{fig:gauge}
\end{figure}

A second example of trivial phase at low temperature is the
3d classical $\mathbb{Z}_2$ lattice gauge theory with an external field \cite{3dgauge_wegner71,3dgauge_phasediagram79}: 
\begin{equation}\label{eq:Hgauge}
 H^{\rm gauge}=\beta h\sum_e z_e
       -\beta \sum_{\text{plaquette }P} Z_P
\end{equation}
where the bits $z_e\in\{0,1\}$ live on edges $e$ of the 3d square lattice, and $P$ sums over all plaquettes containing four neighboring edges in the lattice with $Z_P:=\prod_{e\in P}(1-2z_e)$. From the $\mathbb{Z}_2$ symmetry we focus on the case $h\ge 0$ favoring $0$ states. Since the product of $Z_P$ for six plaquettes belonging to a cube multiplies to identity, excited plaquettes $P$ with $Z_P=-1$ form closed loops in the dual lattice, as shown in Fig.~\ref{fig:gauge}(a). In a background of most bits being $0$, these loop excitations are boundaries of two dimensional membranes of $z_e=1$. One can convert \eqref{eq:Hgauge} to the previous form $H=\sum_u H_u$ where $u$ are vertices of a modified square lattice, but we do not do that explicitly. 

At zero field $h=0$, $H^{\rm gauge}$ has two phases separated by a critical $\beta_{\rm c}^{\rm gauge}$, where $\beta>\beta_{\rm c}^{\rm gauge}$ is a ``deconfined'' phase with classical memory, i.e. the Glauber dynamics mixes slowly.
More precisely, the low-temperature Gibbs state is dominated by sparse small loop excitations of size $\lesssim \log N$, and a constant bits of logical information is dynamically protected by the fact that loop excitations cannot proliferate. See \cite{3dgauge_sharp,3dgauge_highT_unique26} for rigorous statements on the two phases.
Intriguingly, such a classical memory is stable against a small field: the Glauber dynamics is proven to mix slowly if $\beta h\le c_{\rm d}$ and $\beta>\beta_{\rm d}$ for some numerical constants $c_{\rm d},\beta_{\rm d}>0$ \cite{3dgauge_hastings19,3dgauge_slowmix26}. In the phase diagram Fig.~\ref{fig:gauge}(b) at a nonzero $h$, this memory phase only appears at intermediate temperatures, meaning it is driven by entropy effects rather than energy barriers. In fact, this stability generalizes to any classical perturbations \cite{3dgauge_slowmix_general26}. 
We refer to \cite{3dgauge_slowmix26} for a detailed description of the phase diagram from symmetry perspectives together with additional references. 

Here we are interested in the low temperature regime with relatively large field \begin{equation}\label{eq:bh>cH}
    \beta h> c_{\rm H}
\end{equation}
corresponding to the Higgs phase in Fig.~\ref{fig:gauge}(b), where $c_{\rm H}$ is some numerical constant. For any constant $\beta h$, we expect the model violates SSM at sufficiently large $\beta$ (i.e. sufficiently small $h$), as shown in Fig.~\ref{fig:gauge}(c). Nevertheless, we are able to prove fast mixing in this regime, together with the quasi-local connection.

\begin{cor}\label{cor:gauge}
    If \eqref{eq:bh>cH} holds for a numerical constant $c_{\rm H}>0$, the Gibbs state of \eqref{eq:Hgauge} satisfies the same conclusions in Corollary \ref{cor:potts} with different constants $c_1,c_2,\Delta>0$ determined by $h,\beta$.
\end{cor}

\begin{proof}
    This model also satisfies the Peierls condition \eqref{eq:peierls}, because the plaquette terms in \eqref{eq:Hgauge} are already minimized in the all-zero state. It actually holds for the smooth path $H_s=sH^{\rm gauge}+(1-s)\beta h\sum_e z_e$. The proof of Corollary \ref{cor:Ising_frustrated} then applies here because the Peierls condition along such a path is the only model-specific input there.
\end{proof}

\section{Generalizing to the commuting case}\label{sec:commuting}

In this section, we consider general quantum commuting Hamiltonians, and generalize the notion of SDOC together with its consequences. Comparing to the classical case, we will work with the canonical purification \eqref{eq:purification} of Gibbs states that involves the system and its copy. Note that the canonical purification is basis dependent, so we just pick a local basis $\bz$ as the computational basis, which has nothing to do with whether the Hamiltonian is classical. We introduce the following notation: For $S\subset\Lambda$, write $\widetilde{S}$ for its version in the second copy and $\bm{S}$ for the union of the two copies of $S$ in the doubled system. $\bm{S}=\bm{u}$ in the case $S=\{u\}$. The doubled system as two disconnected graphs inherits the graph distance $\mathrm{dist}(\cdot,\cdot)$ of the original graph $G$.

\subsection{Strong decay of correlations}

Consider $A,C\subset \Lambda$ in Definition \ref{def:SDOC}. There is a natural resampling operation that locally resamples the subsystem $C^{\rm c}$ based on the possibly entangled state on $C$, namely the Petz recovery map \cite{Petz88}\begin{align}\label{eq:Petz=}
    \cR_{C^{\rm c}}[\rho] :=& \sqrt{\pi}\lr{\pi_C^{-1/2}\rho_C\pi_C^{-1/2}}\sqrt{\pi} \nonumber\\
    =& \sqrt{\pi_{\overline{C^{\rm c}}}}\lr{\pi_{\partial (C^{\rm c})}^{-1/2}\rho_C\pi_{\partial (C^{\rm c})}^{-1/2}}\sqrt{\pi_{\overline{C^{\rm c}}}} 
\end{align}
acting on any state $\rho$.
It has several properties: \begin{enumerate}
    \item It is a quantum channel, i.e. a completely positive trace-preserving map. 
    \item It is supported in $\overline{C^{\rm c}}$ because the Hamiltonian terms outside that region commute through and cancel to get the second line of \eqref{eq:Petz=}.
    \item It resamples $C^{\rm c}$ in the sense that \eqref{eq:Petz=} is determined by $\rho_C$. In other words one first traces out $C^{\rm c}$ in the operation.
    \item It recovers the global Gibbs state \begin{equation}\label{eq:Rpi=pi}
    \cR_{C^{\rm c}}[\pi_C]=\cR_{C^{\rm c}}[\pi]=\pi.
\end{equation}
    from the marginal Gibbs state in $C$ alone. As a special case, it preserves the global Gibbs state.
    \item The Lindbladian \eqref{eq:glauber=} satisfies $\cL_u=\cR_{\{u\}}-\mathrm{Id}$ (identity channel).
\end{enumerate}
The dual map \begin{equation}
    \cR_{C^{\rm c}}^\dagger[\OO] = \pi_C^{-\frac{1}{2}}\trace_{C^{\rm c}}\lr{\sqrt{\pi}\OO\sqrt{\pi}}\pi_C^{-\frac{1}{2}}
\end{equation}
of $\cR_{C^{\rm c}}$ acts on operators, which as a superoperator is still supported in $\overline{C^{\rm c}}$. 
In fact, the resampling nature makes $\cR_{C^{\rm c}}^\dagger[\OO]$ supported in $\partial C^{\rm c}$ if $\OO$ is supported in $\overline{C^{\rm c}}$.

\begin{defn}[Strong decay of correlations for the commuting case]\label{def:QSDOC}
    Let $f=f(\cdot)$ be a monotonically non-increasing function and $R_{\rm A}\ge 1$. We say that a commuting Gibbs state $\pi$ satisfies $(R_{\rm A},f)$-SDOC if the following holds. For any regions $A,C$ with distance $r$ and $A$ bounded by \eqref{eq:A<cA} as in Definition \ref{def:SDOC}, the Petz recovery almost thermalizes $A$ in the sense that 
\begin{equation}\label{eq:QSDOC}
        \norm{\cR_{C^{\rm c}}^\dagger [\OO]-\alr{\OO}_\pi}
        \le f(r)
    \end{equation}
    for any operator $\OO$ supported in $A$ with $\norm{\OO}=1$.
\end{defn}

To see the meaning of \eqref{eq:QSDOC}, observe that $\cR_{C^{\rm c}}^\dagger [\OO]=\alr{\OO}_\pi$ is equivalent to the statement that the reduced density matrix on $A$ of $\cR_{C^{\rm c}}[\rho]$ for any $\rho$ equals $\pi_A$. The above Definition reduces exactly to the previous Definition \ref{def:SDOC} for the classical case, because $\pi(\cdot|\bz_C)$ is just $\cR_{C^{\rm c}}$ acting on a state with $\bz_C$.

Like Proposition \ref{prop:ratio_cor} for the classical case, one can verify that \eqref{eq:QSDOC} is equivalent to the decay of the ratio-correlation \eqref{eq:ratio_cor} up to an $R_{\rm A}$-dependent prefactor, where the classical ratio should be replaced by the symmetric quantum operator $\lr{\pi_A^{-\frac{1}{2}}\otimes \pi_C^{-\frac{1}{2}}}\pi_{AC}\lr{\pi_A^{-\frac{1}{2}}\otimes \pi_C^{-\frac{1}{2}}}$. In the general quantum case however, it is not obvious whether the two decays based on the recovery map and the ratio-correlation are equivalent. We expect the approximate recovery map from the local Markov property \cite{local_markov} is useful for generalizing the following consequences to general quantum Gibbs states; we leave this extension as future work.

The recovery map has an equivalent formulation in terms of canonical purifications. Define $\kett{\Omega_{C^{\rm c}}}:=\sum_{\bz_{C^{\rm c}}}\ket{\bz_{C^{\rm c}}}\ket{\bz_{C^{\rm c}}}$ in system $\bm{C^{\rm c}}$ and an isometry $\cU_{C^{\rm c}}$ mapping any state $\ket{\phi_C}$ on system $C$ to system $\Lambda \sqcup \widetilde{C^{\rm c}}=\bm{C^{\rm c}}\sqcup C$:
\begin{equation}\label{eq:U_recover}
    \cU_{C^{\rm c}}\ket{\phi_C}:=\sqrt\pi\cdot\pi_{C}^{-1/2}\lr{\kett{\Omega_{C^{\rm c}}}\otimes\ket{\phi_C}}
\end{equation}
where $\pi,\pi_C$ are operators in the first copy $\Lambda$. $\cU_{C^{\rm c}}$ is an isometry $\norm{\cU_{C^{\rm c}}\ket{\phi_C}} = \norm{\ket{\phi_C}}$ because \begin{align}
    \cU_{C^{\rm c}}^\dagger\cU_{C^{\rm c}}
    &=\pi_C^{-1/2}\bbra{\Omega_{C^{\rm c}}}\pi
      \kett{\Omega_{C^{\rm c}}}\pi_C^{-1/2}\nonumber\\
    &=\pi_C^{-1/2}\lr{\trace_{C^{\rm c}}\pi}\pi_C^{-1/2}
      =I_C.
\end{align}
In fact, $\cU_{C^{\rm c}}$ is just a Stinespring dilation of $\cR_{C^{\rm c}}$: \begin{equation}
    \trace_{\widetilde{C^{\rm c}}}\lr{\cU_{C^{\rm c}}\rho_C\cU_{C^{\rm c}}^\dagger} = \cR_{C^{\rm c}}[\rho].
\end{equation}
In particular, $\cU_{C^{\rm c}}$ only acts on the subsystem $\partial C^{\rm c}$ like $\cR_{C^{\rm c}}$. The isometry defines the projector 
\begin{equation}\label{eq:PI=UU}
    \Pi_{C^{\rm c}}:=\cU_{C^{\rm c}}\cdot \cU_{C^{\rm c}}^\dagger
\end{equation}
supported in $\overline{C^{\rm c}}\sqcup \widetilde{C^{\rm c}}$, which is a generalization of \eqref{eq:PI_B} in the classical case that projects onto the local thermal states. One can verify that $\Pi_{C^{\rm c}}$ projects onto states of form \begin{equation}\label{eq:OIOmega}
    \lr{\sqrt{\pi}\OO_{C}\otimes I}\kett{\Omega},
\end{equation}
and satisfies the following properties 
\begin{subequations}\label{eq:Pi_property}
    \begin{align}
\Pi_{C^{\rm c}}\kett{\sqrt\pi}&=\kett{\sqrt\pi}, \label{eq:Pipi=pi}\\
    \Pi_{C^{\rm c}}\Pi_S =\Pi_S\Pi_{C^{\rm c}}&=\Pi_S, \qquad (\text{if } C^{\rm c}\subset S)
    \label{eq:PIPI=PIPI=PI}\\
    \Pi_{C^{\rm c}}\OO\Pi_{C^{\rm c}}&=\cU_{C^{\rm c}}\cdot \cR_{C^{\rm c}}^\dagger[\OO]\cdot \cU_{C^{\rm c}}^\dagger \qquad (\forall \OO \text{ supported in }\Lambda). \label{eq:PIOPI=RO}
\end{align}
\end{subequations}
In particular, \eqref{eq:PIOPI=RO} follows from \eqref{eq:PI=UU} and the Stinespring dilation, which implies \eqref{eq:QSDOC} is equivalent to \begin{equation}\label{eq:PIO-OPI<f}
    \norm{\Pi_{C^{\rm c}}\lr{\OO-\alr{\OO}_\pi}\Pi_{C^{\rm c}}}\le f(r).
\end{equation}


\subsection{Quasi-local connection}
Using the above SDOC condition,
the quasi-local connection Theorem \ref{thm:cohe} generalizes almost verbatim.

\begin{thm}[Strong decay of correlations implies quasi-local connection: the commuting case]\label{thm:cohe_q}
    Theorem \ref{thm:cohe} still holds if the Hamiltonians are commuting quantum ones with \begin{equation}\label{eq:dH_commute_H}
        [\partial_sH_u(s),H(s)]=0 \qquad (\forall u\in \Lambda),
    \end{equation}
    and the coherent Gibbs state connection \eqref{eq:kpi=Ukpi} is replaced by \eqref{eq:pi=Upi} for canonical purifications, where $W_{u,\ell}(s)$ ($\cL^{1\leftarrow 0}_{u,\ell}(s)$) acts within distance $2^\ell$ ($2^\ell+1$) to $\bm{u}$ ($u$) for $\ell<\ell_{\rm max}$.
\end{thm}
\comment{Consider a path of commuting Hamiltonians $H(s)=\sum_{u\in \Lambda} H_{u}(s)$ ($s\in[0,1]$) where each $H_{u}(s)$ is differentiable with \eqref{eq:dsH<1}. If at any $s$ in the path, \eqref{eq:Hu<1} holds for $H=H(s)$ and the Gibbs state $\pi_s$ at $s$ satisfies $(2,f)$-SDOC where function $f$ does not depend on $s$. Then for any integer $\ell_{\rm max}>1$, the following two statements hold: \begin{enumerate}
        \item The canonical purifications are connected by \eqref{eq:pi=Upi} with a Hermitian $W(s)$ given by \eqref{eq:W=suml} with the stated supports and decay \eqref{eq:W_decay}. 

    \item The states are connected in both directions by \eqref{eq:2way}, where the Lindbladian \eqref{eq:L=suml} satisfies the stated quasi-local supports and decay \eqref{eq:L_decay}. The other direction holds as well.
    \end{enumerate}
    }

\begin{proof}
Following the proof of Theorem \ref{thm:cohe}, we suppress the $s$ dependence in notations, and first focus on the canonical purifications. Due to commutativity \eqref{eq:dH_commute_H}, the derivative calculation \eqref{eq:dpi=} generalizes directly to
\begin{align}
    \partial_s\kett{\sqrt\pi}=-\frac12\sum_{u\in \Lambda}(\partial_sH_u-\alr{\partial_sH_u}_\pi)\kett{\sqrt\pi}
    \label{eq:q_derivative}
\end{align}
where operator $H_u$ is acting on the first copy only.
Focusing on a fixed $u$, for any $1\le \ell<\ell_{\rm max}$ we define $\Pi_\ell$ to be the projector \eqref{eq:PI=UU} for $C^{\rm c}$ being the radius-$(2^\ell-1)$ ball centered at $u$. Further defining \begin{equation}\label{eq:q_Pi_ell}
    \Pi_0:=I,\qquad
    \Pi_{\ell_{\rm max}}:=\kett{\sqrt\pi}\bbra{\sqrt\pi},
\end{equation}
we have \begin{align}
    \frac{1}{2}(\partial_sH_u-\alr{\partial_sH_u}_\pi)\kett{\sqrt\pi} &= \frac{1}{2}\lr{\Pi_0 -\Pi_{\ell_{\rm max}}}\partial_sH_u\kett{\sqrt\pi} \nonumber\\
    &= \frac{1}{2}\sum_{\ell=1}^{\ell_{\rm max}} \lr{\Pi_{\ell-1} -\Pi_{\ell}}\partial_sH_u\kett{\sqrt\pi} \nonumber\\
    &= \ii\sum_{\ell=1}^{\ell_{\rm max}} W_{u,\ell} \kett{\sqrt{\pi}} \label{eq:q_H=W}
\end{align}
for Hermitian operators
\begin{equation}
    W_{u,\ell}:=-\frac{\ii}{2}[\Pi_{\ell-1}(\partial_sH_u)\Pi_{\ell-1},\Pi_\ell],\label{eq:q_W}
\end{equation}
with the same supports as those in \eqref{eq:W=suml}.
The last line of \eqref{eq:q_W} follows from \eqref{eq:Pipi=pi} and \eqref{eq:PIPI=PIPI=PI}. Combining with \eqref{eq:q_derivative} and integrating over $s$ then imply \eqref{eq:pi=Upi}. The decay of $W_{u,\ell}$ in \eqref{eq:W_decay} follows from $\norm{W_{u,1}}\le\norm{\partial_sH_u}\le\beta'$ and
\begin{align}
    \norm{W_{u,\ell}}&=\frac12\norm{\mlr{\Pi_{\ell-1}(\partial_sH_u-\alr{\partial_sH_u}_\pi)\Pi_{\ell-1}, \Pi_\ell}}+ \frac{1}{2}\norm{\mlr{\Pi_{\ell-1}\alr{\partial_sH_u}_\pi\Pi_{\ell-1}, \Pi_\ell}}\hspace{15em}\mytag{\text{triangle inequality}}\\
    &\le\norm{\Pi_{\ell-1}(\partial_sH_u-\alr{\partial_sH_u}_\pi)\Pi_{\ell-1}} \mytag{\text{Using \eqref{eq:PIPI=PIPI=PI}}}\\
    &\le \beta'\cdot f(2^{\ell-1}-1)
\end{align}
for $\ell>1$, using \eqref{eq:PIO-OPI<f} equivalent to the SDOC condition.

We then turn to the Lindbladian connection. The derivative \eqref{eq:dspi=} still holds from \eqref{eq:dH_commute_H}, which we write as a symmetrized version \begin{equation}\label{eq:dspi=piHpi}
    \partial_s \pi = -\sqrt{\pi} \sum_{u\in \Lambda} \lr{\partial_sH_u-\alr{\partial_sH_u}_\pi} \sqrt{\pi}.
\end{equation}
For a fixed $u$, we again write  \eqref{eq:=sumVr} where each $V_\ell$ is defined by $\widetilde{V}_\ell$s in \eqref{eq:V=tV}. The difference is that here \begin{equation}\label{eq:tV=R}
    \widetilde{V}_\ell:=\cR_{B_L}^{\dagger}[\partial_sH_u]
\end{equation}
comparing to the classical version \eqref{eq:tV=}, where $B_L$ is still the radius-$2^\ell-1$ ball centered at $u$. Note that here $\widetilde{\cdot}$ does not mean the second copy because there is no second system in the rest of this proof. The operators $V_\ell$ have the same supports ($\overline{B_L}$ except for the $V_{\ell_{\rm max}}$) and bounds \eqref{eq:VLmin}, \eqref{eq:Vr<} as in the previous proof, because the SDOC condition is generalized directly. Similar to \eqref{eq:jump}, we have \begin{equation}\label{eq:piVpi=sigma}
    -\sqrt{\pi}V_\ell\sqrt{\pi} = \norm{V_\ell}\lr{\sigma_\ell-\pi}
\end{equation}
where
\begin{equation}\label{eq:sigmal=}
    \sigma_\ell:=\sqrt\pi\lr{I-\frac{V_\ell}{\norm{V_\ell}}}\sqrt\pi\ge0
\end{equation}
(again, $\sigma_\ell=\pi$ if $V_\ell=0$) is a normalized quantum state because $\tr{V_\ell \pi}=0$. This can be seen from the more general relation \begin{align}
     \tr{\OO\sqrt{\pi} V_\ell\sqrt{\pi}} &= \tr{\OO\sqrt{\pi} \widetilde{V}_{\ell-1}\sqrt{\pi}}-\tr{\OO\sqrt{\pi} \widetilde{V}_\ell\sqrt{\pi}} \mytag{\text{Using \eqref{eq:V=tV}}}\\
     &= \tr{\cR_{B_{L'}}[\sqrt{\pi}\OO\sqrt{\pi}]\partial_s H_u} - \tr{\cR_{B_{L}}[\sqrt{\pi}\OO\sqrt{\pi}]\partial_s H_u} \hspace{10em} \mytag{\text{Using \eqref{eq:tV=R}, } L'=2^{\ell-1}-1}\\
     &=\tr{\sqrt{\pi}\OO\sqrt{\pi}\partial_s H_u} - \tr{\sqrt{\pi}\OO\sqrt{\pi}\partial_s H_u} \mytag{\sqrt{\pi}\OO\sqrt{\pi}\text{ is already thermal in }\overline{B_L}}\\
     &= 0 \label{eq:OpiV=0}
\end{align}
for $1<\ell<\ell_{\rm max}$ and any operator $\OO\ge 0$ supported in $C_\ell:=\lr{\overline{B_L}}^{\rm c}$, where we have used \eqref{eq:Rpi=pi} and the locality of the recovery maps; one can verify that \eqref{eq:OpiV=0} holds for $\ell=1$ and $\ell_{\rm max}$ as well with $C_{\ell_{\rm max}}:=\emptyset$. As a result,
\begin{equation}\label{eq:q_jump}
    \cL^{1\leftarrow0}_{u,\ell}[\rho]
    :=\norm{V_\ell}\mlr{\sqrt{\sigma_\ell}\,\pi_{C_\ell}^{-1/2}
    \rho_{C_\ell}
    \pi_{C_\ell}^{-1/2}\sqrt{\sigma_\ell}-\rho}
\end{equation}
is a valid Lindbladian because
the first term in \eqref{eq:q_jump} is a quantum channel. More precisely, the first term is manifestly completely positive, and trace-preserving: 
\begin{align}
    \tr{\sqrt{\sigma_\ell}\,\pi_{C_\ell}^{-1/2}
    \rho_{C_\ell}
    \pi_{C_\ell}^{-1/2}\sqrt{\sigma_\ell}} &= \tr{
    \rho_{C_\ell}
    \pi_{C_\ell}^{-1}\pi} - \frac{1}{\norm{V_\ell}} \tr{\lr{\pi_{C_\ell}^{-1/2}\rho_{C_\ell}\pi_{C_\ell}^{-1/2}}\sqrt{\pi} V_\ell \sqrt{\pi}} \hspace{15em}\mytag{\text{Using \eqref{eq:sigmal=} and cyclicity}}\\
    &=\trace_{C_\ell}\lr{
    \rho_{C_\ell} 
    \pi_{C_\ell}^{-1}
    \pi_{C_\ell}} - 0\mytag{\text{Using \eqref{eq:OpiV=0}}}\\
    &= \tr{\rho}.
\end{align}

Furthermore, \eqref{eq:q_jump} is supported in $\overline{\overline{B_L}}$ (comparing to $\overline{B_L}$ for the previous classical case). To see this,
observe that \begin{equation}\label{eq:sigmal='}
    \sigma_\ell = \sigma'_\ell \cdot \ee^{-H_{\ell}^{\rm out}}
\end{equation}
where $H_{\ell}^{\rm out}$ is the Hamiltonian terms supported in $C_\ell$ and $\sigma'_\ell$ is supported in $\overline{\overline{B_L}}$. Crucially, the two operators on the right hand side of \eqref{eq:sigmal='} commute with each other: $H_{\ell}^{\rm out}$ commutes with $V_\ell$ due to disjoint support, and commutes with $H-H_{\ell}^{\rm out}$ that gives rise to $\sigma'_\ell$. As a result, \begin{equation}
    \sqrt{\sigma_\ell} = \sqrt{\sigma'_\ell} \cdot \ee^{-\frac{1}{2}H_{\ell}^{\rm out}}
\end{equation}
and the second factor cancels precisely the $H^{\rm out}_\ell$ contributions in $\pi_{C_\ell}^{-1/2}$ in \eqref{eq:q_jump}, leading to the $\overline{\overline{B_L}}$ support of superoperator \eqref{eq:q_jump}.

Combining \eqref{eq:dspi=piHpi}, \eqref{eq:=sumVr}, \eqref{eq:piVpi=sigma} and \eqref{eq:q_jump} then leads to \begin{equation}
    \partial_s \pi = \sum_{u\in \Lambda}\sum_{\ell=1}^{\ell_{\rm max}} \cL^{1\leftarrow0}_{u,\ell}[\pi],
\end{equation} 
which integrates to \eqref{eq:2way}.
The quasi-local decay \eqref{eq:L_decay} follows from the same $\norm{V_\ell}$ bounds as the previous proof, and the other direction $0\leftarrow 1$ also holds by symmetry.
\end{proof}

\subsection{Fast mixing}
The spectral gap Theorem \ref{thm:gap} also generalizes.
\begin{thm}\label{thm:gap_q}
    Theorem \ref{thm:gap} still holds for commuting Gibbs state $\pi$ and the gap of its Glauber dynamics \eqref{eq:glauber=}, if the condition $\kpi=U\ket{\bm{+}}$ is replaced by 
\begin{equation}\label{eq:q_U_product}
    \kett{\sqrt\pi}=U\frac{\kett{\Omega}}{\norm{\kett{\Omega}}},
\end{equation}
    and the Lieb--Robinson bound \eqref{eq:LRB} is understood on the doubled system viewing $\bm{u}$ as one site. 
\end{thm}

A similar corollary like Corollary \ref{cor:SDOC_gap} follows from combining with Theorem \ref{thm:cohe_q}, which we omit here for brevity. We also remark that while we consider the specific heat-bath Lindbladian \eqref{eq:Lu=}, the same result generalizes to Davies generators \cite{davies1974} built from a complete basis of single-site operators and a uniform $\Theta(1)$ jump rate.

\begin{proof}
We follow the proof of Theorem \ref{thm:gap}. 

First, the ``local equilibrium'' projector $\Pi_u$ is defined to be \eqref{eq:PI=UU} with 
$C^{\rm c}=B_u$, i.e. the ball of radius $R-1$ centered at $u$.
$\Pi_u$ is supported in $\overline{B_u}$, and satisfies the properties \eqref{eq:Pi_property}. We still define the dressed $\widetilde{\Pi}_u$ by \eqref{eq:tildePI}. Note that only for this notation $\widetilde{\cdot}$ does not mean the second copy.

We then establish the Poincare inequality from somewhat direct computations, whose proof is given later.
\begin{lem}\label{lem:poincare_q}
    For any $u$, \begin{equation}\label{eq:Poincare_q}
    \sum_{v\in \overline{B_u}} \cH_v \ge \ee^{-2\beta|\overline{\overline{B_u}}|} (I-\Pi_u).
\end{equation}
\end{lem}
Comparing to \eqref{eq:Poincare}, \eqref{eq:Poincare_q} includes sites $v\in \partial B_u$.
Following \eqref{eq:UHU>DeltaR}, we have 
\begin{align}
    \bbra{\psi}U^\dagger \cH U\kett{\psi} &\ge \bbra{\psi}U^\dagger \lr{\frac{1}{c_dR^d} \sum_{u\in \Lambda}\sum_{v\in \overline{B_u}} \cH_{v} }U \kett{\psi} \nonumber\\
    &\ge \frac{\ee^{-2\cdot 3^dc_d\beta R^d}}{c_dR^d} \sum_{u\in \Lambda} \bbra{\psi}(I-\widetilde{\Pi}_u)\kett{\psi} 
    \label{eq:UHU>DeltaR_q}
\end{align}
using $|\overline{B_u}|\le c_d R^d$ and $|\overline{\overline{B_u}}|\le c_d(R+1)^d\le 3^dc_dR^d$.
This is the only step where the Poincare inequality is used.

Define
\begin{equation}\label{eq:Q_quantum}
    Q_v^\perp := \frac{1}{\norm{\kett{\Omega_v}}^2} \kett{\Omega_v}\bbra{\Omega_v},\qquad Q_v = I-Q_v^\perp.
\end{equation}
We then verify the dressed-projector estimate \eqref{eq:sumr<}, which boils down to bounding the two terms in \eqref{eq:sumr<+}. The second term in \eqref{eq:sumr<+} is bounded by Lieb-Robinson bound alone, so one can verify that \eqref{eq:sumr>_2ndterm} generalizes directly. For the first term in \eqref{eq:sumr<+},
\eqref{eq:QPI=} still holds with $\ket{\psi}$ replaced by $\kett{\psi}$ satisfying \begin{equation}\label{eq:PIkett=kett}
    \Pi_u\kett{\psi} = \kett{\psi},
\end{equation}
and Lieb-Robinson bound yields an operator $\OO'$ supported in $\bm{A}$ satisfying \eqref{eq:LRB'}, where $A\subset\Lambda$ is the radius-$R_1$ ball centered at $u$. So we reproduce \eqref{eq:QPI<'} by  
\begin{align}
    \norm{Q_u\widetilde{\Pi}_u }^2 &\le \max_{\psi} \bbra{\psi}\OO'\kett{\psi} + \epsilon \mytag{\text{Using \eqref{eq:LRB'}}}\\
    &\le \bbra{\sqrt{\pi}}\OO'\kett{\sqrt{\pi}} + \max_{\psi}\abs{\bbra{\psi}\OO'\kett{\psi}-\bbra{\sqrt{\pi}}\OO'\kett{\sqrt{\pi}}}+ \epsilon \nonumber\\
    &\le \mlr{\bbra{\sqrt{\pi}}UQ_u U^\dagger\kett{\sqrt{\pi}}+ \epsilon} + \norm{\OO'}\cdot 2f(R-R_1-1) + \epsilon \hspace{20em}\mytag{\text{Using \eqref{eq:LRB'} and SDOC}} \\
    &\le \norm{\kett{\Omega}}^{-2}\bbra{\Omega}Q_u\kett{\Omega} + 2\epsilon + 2(1+\epsilon)f(R-R_1-1) \mytag{\text{Using \eqref{eq:LRB'} and }\kpi=U\kett{\Omega}/\norm{\kett{\Omega}}}\\
    &= 2\epsilon + 2(1+\epsilon)f(R-R_1-1) \label{eq:QPI<'_q}
\end{align}
Here SDOC is applied in the third line as follows. Although $\OO'$ acts in the doubled system, we can use isometry \eqref{eq:U_recover} to get operator 
\begin{equation}\label{eq:O''=O'}
    \OO'':=\cU_A^\dagger\OO'\cU_A
\end{equation}
supported in $\partial A$ in the first copy. For any state \eqref{eq:PIkett=kett} of the form \eqref{eq:OIOmega} for $C=B_u^{\rm c}$, \begin{equation}\label{eq:psi=OC}
    \kett{\psi}=\cU_A(\sqrt{\pi_{A^{\rm c}}}\OO_C\otimes I)\kett{\Omega_{A^{\rm c}}},
\end{equation} 
which leads to \begin{align}
    \bbra{\psi}\OO'\kett{\psi}&= \bbra{\Omega_{A^{\rm c}}}(\sqrt{\pi_{A^{\rm c}}}\OO_C\otimes I)^\dagger\OO''(\sqrt{\pi_{A^{\rm c}}}\OO_C\otimes I)\kett{\Omega_{A^{\rm c}}} \mytag{\text{Using \eqref{eq:O''=O'}}}\\
    &= \trace_{A^{\rm c}}\lr{\OO'' \sqrt{\pi_{A^{\rm c}}}\OO_C\OO_C^\dagger\sqrt{\pi_{A^{\rm c}}}} \nonumber\\
    &= \trace_{A^{\rm c}}\mlr{\OO'' \trace_A\lr{\sqrt{\pi}\OO_C\OO_C^\dagger\sqrt{\pi}} } \mytag{H \text{ is commuting}}\\
     &= \tr{\OO'' \sqrt{\pi}\OO_C\OO_C^\dagger\sqrt{\pi} } \nonumber\\
     &=\bbra{\psi}\OO''\kett{\psi} \label{eq:O'=O''}
\end{align}
using \eqref{eq:OIOmega}. \eqref{eq:O'=O''} also holds for the special case $\kett{\psi}=\kett{\sqrt{\pi}}$. As a result, SDOC implies \begin{align}
    \abs{\bbra{\psi}\OO'\kett{\psi}-\bbra{\sqrt\pi}\OO'\kett{\sqrt\pi}} &=\abs{\bbra{\psi}\OO''\kett{\psi}-\bbra{\sqrt\pi}\OO''\kett{\sqrt\pi}} \mytag{\text{Using \eqref{eq:O'=O''}}}\\
    &\le\norm{\Pi_u\lr{\OO''-\alr{\OO''}_\pi}\Pi_u}
      \mytag{\text{Using \eqref{eq:PIkett=kett}}}\\
    &\le\norm{\OO'}f(R-R_1-1)
      \label{eq:q_QPi_bound}
\end{align}
using 
$\norm{\OO''}\le\norm{\OO'}$, which establishes \eqref{eq:QPI<'_q}.

We have established the two key Lemmas \ref{lem:poincare_q} and \ref{lem:H>X}. The rest of the proof of Theorem \ref{thm:gap} then directly follows by the proof of Theorem \ref{thm:gap}, where because the constant prefactor in \eqref{eq:UHU>DeltaR_q} is tigher here the same bound \eqref{eq:DeltaH>} remains valid.

\end{proof}

Finally, we prove the Poincare inequality in the commuting case. 
\begin{proof}[Proof of Lemma \ref{lem:poincare_q}]
For the Glauber Lindbladian \eqref{eq:glauber=}, the corresponding Hamiltonian \eqref{eq:H=piLpi} acting on the doubled Hilbert space is
\begin{align}\label{eq:q_heatbath_H}
    \cH_u &=I- \mathcal{D}[\pi^{\frac{1}{4}}_{\overline{u}}]\cdot \lr{\kett{\Omega_u}\bbra{\Omega_u}\otimes \mathcal{D}[\pi^{-\frac{1}{2}}_{\partial u}]} \cdot \mathcal{D}[\pi^{\frac{1}{4}}_{\overline{u}}]\nonumber\\
    &=I- \mathcal{D}[\pi^{\frac{1}{4}}]\cdot \lr{\kett{\Omega_u}\bbra{\Omega_u}\otimes \mathcal{D}[\pi^{-\frac{1}{2}}_{u^{\rm c}}]} \cdot \mathcal{D}[\pi^{\frac{1}{4}}]
\end{align}
where \begin{equation}
    \DD[\OO]:=\OO\otimes \OO^{\rm T}
\end{equation}
is the doubled operator of $\OO$ with $\OO^{\rm T}$ being its transpose in the computational basis.
\eqref{eq:q_heatbath_H} satisfies \begin{equation}\label{eq:Hu>I-PIu}
    \cH_u\ge I-\widehat\Pi_{\{u\}}
\end{equation}
where $\widehat\Pi_S$ is the projector onto states of form \begin{equation}\label{eq:PIhat_states}
   \DD[\pi^{\frac{1}{4}}] (\OO_{S^{\rm c}}\otimes I)\kett{\Omega} =  (\pi^{\frac{1}{4}}\OO_{S^{\rm c}}\pi^{\frac{1}{4}}\otimes I)\kett{\Omega}
\end{equation}
for any operator $\OO_{S^{\rm c}}$ supported in $S^{\rm c}$.
Since $0\le \cH_u\le I$, to check that inequality \eqref{eq:Hu>I-PIu} holds, it suffices to show that any state $(I-\cH_u)\kett{\psi'}$ belongs to \eqref{eq:PIhat_states} for $S=\{u\}$. This is indeed the case from direct computation \begin{align}
    (I-\cH_u)\kett{\psi'} &=\mathcal{D}[\pi^{\frac{1}{4}}]\cdot \lr{\kett{\Omega_u}\bbra{\Omega_u}\otimes \mathcal{D}[\pi^{-\frac{1}{2}}_{u^{\rm c}}]} \kett{\psi} \mytag{\kett{\psi}:= \mathcal{D}[\pi^{\frac{1}{4}}]\kett{\psi'}}\\
    &= \mathcal{D}[\pi^{\frac{1}{4}}]\cdot \lr{\kett{\Omega_u}\bbra{\Omega_u}\otimes \mathcal{D}[\pi^{-\frac{1}{2}}_{u^{\rm c}}]} (\OO\otimes I)\kett{\Omega}\nonumber\\
    &=\mathcal{D}[\pi^{\frac{1}{4}}]\cdot \mlr{\lr{\pi^{-\frac{1}{2}}_{u^{\rm c}}\cdot \trace_u(\OO)\cdot \pi^{-\frac{1}{2}}_{u^{\rm c}}} \otimes I} \kett{\Omega}
\end{align}
where the second line uses the fact that any state can be written as $(\OO\otimes I)\kett{\Omega}$ for some operator $\OO$ acting on the first copy, because operators acting on the second copy can be moved to the first copy for the maximally entangled state $\kett{\Omega}$.

More generally, the projector $\widehat{\Pi}_S$ is related to the previous projector $\Pi_u$ by \begin{equation}\label{eq:hat<Pi}
    \widehat{\Pi}_{\overline{B_u}}\le \Pi_u
\end{equation}
because any state \eqref{eq:PIhat_states} with $S=\overline{B_u}$ can be rewritten as \begin{equation}
    \mlr{\sqrt{\pi}\lr{\pi^{-\frac{1}{4}}\OO_{\lr{\overline{B_u}}^{\rm c }} \pi^{\frac{1}{4}}}\otimes I} \kett{\Omega}
\end{equation}
which is contained in states \eqref{eq:OIOmega} with $C^{\rm c}=B_u$ from the commuting property of $\pi$. 

On the other hand, we relate $\widehat{\Pi}_S$ to projectors \eqref{eq:Q_quantum}.
Observe that by rewriting \eqref{eq:PIhat_states} in the form $\DD[\ee^{-\frac{1}{4}H_{\overline{S}}}](\OO'_{S^{\rm c}}\otimes I)\kett{\Omega}$, $I-\widehat{\Pi}_S$ annihilates any state of form $\DD[\ee^{-\frac{1}{4}H_{\overline{S}}}]\prod_{v\in S}Q_v^\perp \kett{\psi}$, where $H_{\overline{S}}:=\sum_{u\in \overline{S}}H_u$. This implies for any $\kett{\psi}$, we can decompose $\kett{\psi}=\prod_{v\in S}Q_v^\perp\kett{\psi}+\lr{I-\prod_{v\in S}Q_v^\perp}\kett{\psi}$ so that \begin{equation}\label{eq:Q<PI<Q}
    \norm{\lr{I-\widehat{\Pi}_S}\DD[\ee^{-\frac{1}{4}H_{\overline{S}}}]\kett{\psi}}^2 = \norm{\lr{I-\widehat{\Pi}_S}\DD[\ee^{-\frac{1}{4}H_{\overline{S}}}]\lr{I-\prod_{v\in S}Q_v^\perp}\kett{\psi}}^2 \le \ee^{\beta|\overline{S}|}\norm{\lr{I-\prod_{v\in S}Q_v^\perp}\kett{\psi}}^2
\end{equation}
where we have used \begin{equation}\label{eq:eH<eS}
    \ee^{-\frac{1}{4}\beta |\overline{S}|}I\le \ee^{-\frac{1}{4}H_{\overline{S}}} \le \ee^{\frac{1}{4}\beta |\overline{S}|}I.
\end{equation}
from \eqref{eq:Hu<1}. Furthermore, the above relation between $\widehat{\Pi}_S$ and $Q_v$ implies for any $v\in S$,
\begin{align}
    \norm{Q_v\kett{\psi}}
    &\le
    \norm{
        \DD[\ee^{\frac14 H_{\overline{\{v\}}}}]
        (I-\widehat\Pi_{\{v\}})
        \DD[\ee^{-\frac14 H_{\overline{\{v\}}}}]
        \kett{\psi}
    }
    \mytag{
        Q_v\DD[\ee^{\frac14 H_{\overline{\{v\}}}}]
        \widehat\Pi_{\{v\}}=0
    }\\
    &=
    \norm{
        \DD[\ee^{\frac14 H_{\overline S}}]
        (I-\widehat\Pi_{\{v\}})
        \DD[\ee^{-\frac14 H_{\overline S}}]
        \kett{\psi}
    }\hspace{15em}
    \mytag{
        [\widehat\Pi_{\{v\}},
        H_{\overline S}-H_{\overline{\{v\}}}]=0
        \text{ on both copies}
    }\\
    &\le
    \ee^{\frac12\beta|\overline S|}
    \norm{
        (I-\widehat\Pi_{\{v\}})
        \DD[\ee^{-\frac14 H_{\overline S}}]
        \kett{\psi}
    }.
    \label{eq:Q<PI1}
\end{align}
where we have used \eqref{eq:eH<eS} again.

As a result, for any $\kett{\psi'}=\DD[\ee^{-\frac{1}{4}H_{\overline{\overline{B_u}}}}]\kett{\psi}$ we have 
\begin{align}
    \bbra{\psi'}\lr{\sum_{v\in \overline{B_u}} \cH_v}\kett{\psi'} &\ge \sum_{v\in \overline{B_u}}\norm{ \lr{I-\widehat{\Pi}_{\{v\} }} \kett{\psi'}}^2 \hspace{15em}\mytag{\text{Using \eqref{eq:Hu>I-PIu} and projector property}} \\
    &\ge\ee^{-\beta|\overline{\overline{B_u}}|}\sum_{v\in \overline{B_u}}\norm{Q_v\kett{\psi}}^2
       \mytag{\text{Using \eqref{eq:Q<PI1}}}\\
    &\ge \ee^{-\beta|\overline{\overline{B_u}}|}\norm{\lr{I-\prod_{v\in \overline{B_u}}Q_v^\perp}\kett{\psi}}^2
       \mytag{\text{commuting on-site projectors}}\\
    &\ge\ee^{-2\beta |\overline{\overline{B_u}}|}\norm{(I-\widehat\Pi_{\overline{B_u}})\DD[\ee^{-\frac{1}{4}H_{\overline{\overline{B_u}}}}]\kett{\psi}}^2 \mytag{\text{Using \eqref{eq:Q<PI<Q}}}\\
    &\ge \ee^{-2\beta|\overline{\overline{B_u}}|} \bbra{\psi'}(I-\Pi_u)\kett{\psi'}
       \label{eq:q_Poincare_compare}
\end{align}
using \eqref{eq:hat<Pi} at last. This proves \eqref{eq:Poincare_q}.

\end{proof}

\section{Discussion and Outlook}\label{sec:outlook}

Our work opens up several questions regarding general statements of fast mixing:
Our results establish a constant spectral gap for Glauber dynamics, but leave open whether the mixing time can be improved to $\order(\log N)$, i.e. rapid mixing. Notably, the unitary connection idea is also utilized in a recent work \cite{rapidmix_1d_Chen26} to improve rapid mixing results in 1d. Our results do not apply to arbitrary fixed boundary conditions because boundary order can make the dynamics slow mixing. Although SSM may be necessary for a constant spectral gap in this case, there may be a sense in which observables sufficiently far from the boundary equilibrate rapidly even in the presence of boundary order; it is interesting to investigate whether our SDOC conditions could imply such bulk mixing results.
Furthermore, our fast-mixing criterion assumes SDOC along a path connecting the Gibbs state to a product state, a condition satisfied in all applications considered, but it remains open whether SDOC for a single Gibbs state already guarantees fast or rapid mixing.

Beyond the canonical statistical-mechanical models we considered, our results may find applications more broadly. In a companion work \cite{noredundancy26}, we prove SDOC for any quantum stabilizer code without local relations among the code checks at any constant temperature, ruling out proposals for low-dimensional thermal quantum memories. More generally, it is important to understand how to prove SDOC, and how to generalize the condition beyond trivial phases or commuting models. If some generalized version of SDOC holds throughout each thermal phase in known examples, that would establish the $\order(1)$-time quasi-local connections as an universal characterization of thermal phases of matter, in analogy to quasi-local unitary connections for gapped ground-state phases.

For the frustration-free quantum Hamiltonian associated to the Glauber dynamics, our spectral gap proof extends beyond the perturbative neighborhood of a product state. Since existing proofs of ground-state stability are mostly perturbative \cite{stab_yarotsky06,TQO10,michalakis2013stability,ldpc_gap25,ldpc_gap_25,stab_gapless26}, it would be interesting to determine whether this approach can help establish stability of ground-state phases in a broader class of quantum systems.

\section*{Acknowledgements}
We would like to thank Shengqi Sang for useful discussions, and Yaodong Li and Benedikt Placke for collaboration on a companion work \cite{noredundancy26}.  C.Y. was partially supported by the US Department of Energy, Office of Science under Award No. DE-SC0019380 and by the Stanford Q-FARM Bloch Postdoctoral Fellowship in Quantum Science and Engineering. VK acknowledges support from the Packard Fellowship in Science and Engineering and the Office of Naval Research Young Investigator Program (ONR YIP) under Award Number N00014-24-1-2098.

The authors conceived this project and formulated its key conceptual ideas in the context of our companion work \cite{noredundancy26}.  ChatGPT was used to assist in developing the proofs and in writing the manuscript. The authors take full responsibility for the correctness of the paper.

\textbf{Note added.}---
We recently became aware of related work by Blanca et al. \cite{Potts_new26}, which also establishes fast mixing of Glauber dynamics for the Potts model throughout the high-temperature phase, using different techniques than ours. In our time convention, their inverse spectral gap is bounded by $N^{\mathrm{o}(1)}$, whereas Corollary \ref{cor:potts} gives a bound independent of $N$. 

\bibliography{biblio}

\end{document}